\documentclass[5p,times,twocolumn]{elsarticle} 
\usepackage[numbers]{natbib}
\usepackage{multirow}
\usepackage{blindtext, graphicx}
\usepackage{url}
\usepackage{listings}
\usepackage{hyperref}
\usepackage{algorithm}
\usepackage{algorithmic}
\usepackage{amsmath}
\usepackage{booktabs}
\usepackage{caption}
\usepackage{subcaption}
\usepackage{amsfonts}
\usepackage[dvipsnames]{xcolor} 
\usepackage[utf8]{inputenc}
\usepackage{multirow}
\usepackage{ragged2e}
\usepackage{pgfplots}
\usepackage{fancyhdr}
\usepackage{graphicx}
\usepackage[font=small]{caption}
\usepackage[utf8]{inputenc}

\newtheorem{Observation}{Observation}
\newtheorem{Definition}{Definition}

\journal{Information and Software Technology} 

\begin{document}

\let\WriteBookmarks\relax
\def\floatpagepagefraction{1}
\def\textpagefraction{.001}

\title {Investigating The Smells of LLM-generated Code\tnoteref{t1}}

\tnotetext[t1]{This paper is an extended and revised version of the conference paper entitled "\emph{Does LLM Generated Code Smell}?" published by IEEE in the \emph{Proceeding of The 9th International Conference on Cloud, Big Data Computing} (ICCBDC 2025), 14 Aug. 2025, pp. 68-73 \cite{paul2025does}. The results presented in the this paper supersede the conference paper.}

\author{Debalina Ghosh Paul}
\ead{19217422@brookes.ac.uk}
                
\author{Hong Zhu\corref{cor1}} 
\ead{hzhu@brookes.ac.uk}

\author{Ian Bayley}
\ead{ibayley@brookes.ac.uk}

\cortext[cor1]{Corresponding author}

\affiliation {
  organization={School of Engineering, Computing and Mathematics, Oxford Brookes University},
  city={Oxford},
  postcode={OX3 0BP},
  country={UK}}
%
   
\begin{abstract}

\noindent\emph{Context}:

Large Language Models (LLMs) are increasingly being used to generate program code. Much research has been reported on the functional correctness of generated code, but there is far less on code quality. 

\noindent\emph{Objectives}: 

In this study, we propose a scenario-based method of evaluating the quality of LLM-generated code to identify the weakest scenarios, for which the quality of LLM-generated code should be improved. 

\noindent\emph{Methods}: 

The method measures code smells, an important indicator of code quality, and compares them with a baseline formed from reference solutions of professionally written code. The test dataset is divided into various subsets according to the topics of the code and complexity of the coding tasks to represent different scenarios of using LLMs for code generation. We will also present an automated test system for this purpose and report experiments with the Java programs generated in response to prompts given to four state-of-the-art LLMs: Gemini Pro, ChatGPT, Codex, and Falcon.  

\noindent\emph{Results}: 

We find that LLM-generated code has a higher incidence of code smells compared to reference solutions. Falcon performed the least badly, with a smell increase of 42.28\%, followed by  Gemini Pro (62.07\%), ChatGPT (65.05\%) and finally Codex (84.97\%). The average smell increase across all LLMs was 63.34\%, comprising 73.35\% for implementation smells and 21.42\% for design smells.  We also found that the increase in code smells is greater for more complex coding tasks and for more advanced topics, such as those involving object-orientated concepts.  

\noindent\emph{Conclusion}: 

In terms of code smells, LLMs' performances on various coding task complexities and topics are highly correlated to the quality of human-written  code in the corresponding scenarios. However, the quality of LLM-generated code is noticeably poorer than human-written  code. 


\end{abstract}




\lhead{Version 5.0; \today.}
\rhead{D. G. Paul, H. Zhu and I. Bayley}
\thispagestyle{fancy}
\pagestyle{fancy}  
\maketitle
  
\section{Introduction}\label{sec:Introduction}

\subsection{Motivation}

Large language models (LLMs) are increasingly being used in practice to assist programmers with code generation. It is widely recognised that such machine learning (ML) models can significantly improve productivity \cite{Shani2024},
but there are concerns about the quality of the code generated. For example, the \emph{2023 Stack Overflow Developer Survey} conducted by Google in 2023 with over 90,000 respondents globally found that ``\emph{39\% of developers say they don't trust AI-generated
code}'' \cite{DORA2024}.   More recently, based on a quantitative study of 211 million lines of code changes submitted to GitHub between January 2020 and December 2024, Harding et al. \cite{Harding2025AICopilotQuality} identified multiple ``\emph{signs of
eroding code quality}'' such as duplicated code and various forms of technical debt. Moreover, they were able to link this rising defect rate with AI adoption. So this raises an important question: what specific quality defects are present in LLM-generated code? 

Our previous work \cite{ghosh2024} has looked at two of these attributes: correctness and complexity, and evaluated ChatGPT on the benchmark ScenEval that we constructed, where correctness is determined by passing all test cases automatically generated from both ChatGPT generated code and the reference solution, and complexity was measured with cyclomatic complexity, cognitive complexity, and line counts. It was found that correctness was lower for more advanced coding topics and more complex coding tasks. ChatGPT-generated code was more complex than human-written code. In addition, the complexity increases greater for more complex tasks than simpler ones. These suggest that LLMs are less useful for more complex and advanced tasks and that could explain why senior developers use them less often \cite{zdravkovic2025amd},   \cite{deniz2023unleashing}. 

However, Ziegler et al., in their study of GitHub Copilot, observed that the major driving force for the adoption of generated code is not its correctness but whether it is useful as a starting point for further development \cite{ziegler2024measuring}. So, in this paper we will shift our attention to the quality attributes that are relevant to that. These include readability, testability, maintainability, ease of modification and evolution, reusability and so on.

\subsection{Challenges And Our Approach}\label{sec:ChallengesAndApproach}

It is difficult to measure LLM-generated code on these quality attributes since the context of their usage is unknown. Our solution is to use code smell detection techniques since they are well established in software engineering research for this purpose \cite{beck1999bad, suryanarayana2014refactoring, sharma2016configuration}. It is widely recognised that code smells are indicators of problems present in program code with maintenance, evolution and reuse \cite{lacerda2020code}. 

A problem with the concept of code smells, however, is that it is relatively subjective. Beck and Fowler define  that bad code smells are ``\emph{not precise criteria for flaws in program code}'' \cite{beck1999bad,fowler2018refactoring}. They suggest that the presence of smells is ``\emph{better to be judged based on informed human intuition}''. However, research has found that ``\emph{human agreement on smell detection is low}'' \cite{santos2018systematic}. Our solution is to follow best practice in ML research: benchmark LLM performance on a dataset and compare with a baseline. The dataset we will use is ScenEval \cite{ghosh2024}. It consists of tasks collected from both textbooks and questions submitted to StackOverflow; the latter is a professional coding problem-solving website so the questions are real-world. Each task is accompanied by a reference solution either written by the textbook authors or supplied by professional programmers in the IT industry in response to the question on StackOverflow and
 scored highly by peers. These reference solutions provide a good baseline that reflects the current state-of-the-art in professionally written code so that we can decide whether the LLM-generated code is of comparable quality. 

To achieve our research goal, it is insufficient to score each LLM with a single scalar value since they are used in many different contexts for different purposes by different users. So we evaluate the LLMs on different scenarios. These include different problem topics and different complexities. The information we need for filtering on these scenarios is included as metadata with each coding task in the ScenEval benchmark \cite{ghosh2024}. Different subsets of the dataset can therefore be formed easily to represent different scenarios.

However, the existence of these subsets necessitates repeated experiments, so we automate both the execution of the experiments and the subsequent analysis of the large volume of data that is produced. We design and implement an automated test system following the datamorphic testing methodology \cite{zhu2019a} and execute the
experiment with the Morphy test automation environment \cite{zhu2020}. 

\subsection{Contributions}

Our main contributions are as follows. 

\begin{enumerate}
  \item We propose a scenario-based method to investigate the quality of LLM-generated code by detecting code smells, statistically analysing them and comparing them with a baseline of human-written programs. This method enables us to identify the quality weaknesses of LLM-generated code specific to each scenario.
  
  \item We have designed and implemented a test system to automate the experiments with LLMs and the analysis of the data obtained from the experiments. 
  
  \item We have conducted a systematic and intensive experiment with four current state-of-the-art LLMs and compared the generated code with human-written reference solutions in the benchmark as the baseline. The experiment demonstrates the validity and feasibility of the proposed method, and the efficiency and effectiveness of the test system. 
  
  \item We have identified the weaknesses of LLMs for code generation with regards to the quality of the code generated. These results are the first of their kind in the literature as far as we know. 
\end{enumerate}

%

\subsection{Structure of the Paper}

The paper is organised as follows. 
Section \ref{sec:RelatedWork} reviews related work on how to evaluate the quality of LLM-generated code and formulates the research questions.
Section \ref{sec:Background} gives a brief introduction to the notion of code smell and techniques for code smell
detection. Our uses of these techniques are described. 
Section \ref{sec:TestSystem} presents the automated test system, which is designed and implemented based on the datamorphic software testing methodology. 
Section \ref{sec:ExperimentDesign} presents the design of our experiment. 
Section \ref{sec:Results} reports the results and presents the analysis of the data.
Section \ref{sec:ThreatsToValidity} discusses the threats to experimental validity.
Finally, Section \ref{sec:Conclusion} summarises the findings and discusses directions for future work.  

\section{Related Works And Open Problems}\label{sec:RelatedWork}

Evaluation of the capability of LLMs in code generation has hitherto focussed on functional correctness, which is also referred to as code hallucinations by some authors \cite{liu2026beyond},  but far less on code quality and only recently; see \cite{ghosh2024b} for a recent review. We now discuss the few works in the literature that are relevant, followed by the open research questions addressed in this paper. 

\subsection{Manual Evaluation}

In 2024, Miah and Zhu proposed a user-centric methodology to evaluate LLMs according to the quality of code generated \cite{miah2024user}. The method consists of the following three components:
\begin{enumerate}
\item A \emph{multi-attempt} testing process model: the tester engages in an iterative process of interactions with an LLM by (a) formulating, revising and submitting a query to the LLM under test, (b) getting responses from the LLM, (c) assessing the LLM-generated solution for usability, d) determining whether a further attempt of querying the LLM should be made. This iterative process continues until either a satisfactory solution is obtained or a threshold maximum number of allowed iterations is reached. 
\item A set of eight quality attributes related to how easily a human user could use LLM-generated code. These are accuracy, completeness, conciseness, clarity of logic, readability, well-structuredness, parameter coverage and depth of explanation. 
\item A set of three metrics that measure the user experience. The first is the average number of attempts, each of which is an iteration of the human-LLM interaction described above. The second is the average completion time for the task of coding using the LLM. The third is the success rate, where success means that useful code has been generated. 
\end{enumerate}

The authors illustrated the methodology using ChatGPT with 100 tasks in the programming language R. Usability was high: 3.8 out of 5, manually assessed on a Likert scale of 1 to 5 for each of the eight quality attributes. The average number of attempts was only 1.61 and the average completion time was 42 seconds. 

Although subjective manual evaluation is valuable for the user-centred process proposed in the paper \cite{miah2024user}, it is labour-intensive and error-prone. Therefore, objective automated methods are preferable.

\subsection{Automated Evaluation}

As far as we know, the only way to use automation to evaluate code quality is via code smell detection. Siddiq et al. was perhaps the first to do this, in \cite{siddiq2022empirical}, where they used  Pylint \cite{thenault2001pylint} to detect the code smells in three different training datasets: CodeXGlue \cite{lu2021codexglue}, APPS \cite{hendrycks2021measuring}, and Code Clippy \cite{coooper2021code}; Bandit \cite{bandit} was also used in order to detect security code smells. To investigate the impact of code smell in training dataset, ten different code models, each based on the GPT-Neo 125M model, were trained on these three datasets, then tested on the HumanEval dataset \cite{chen2021evaluating} and compared with GitHub Copilot. 

They found that the most frequent smells in the training datasets were also the most frequent in the generated code. They concluded that smells in the training datasets leaked into to the code models, although there was no statistical analysis of the correlations between the two, nor any causality analysis. However, their work has raised concerns about code smells in training datasets. 

Moratis et al. applied code smell detection to the dataset DevGPT of reported iterative conversations in GitHub between developers and ChatGPT ~\cite{moratis2024write}. Two types of conversations were extracted:
\begin{enumerate}
  \item \emph{Write me this code}, with text instructions as input to produce a program code
  \item \emph{Improve this code}, with code snippets as input to improve the quality of the input program code
\end{enumerate} 

The conversations were then fed into the code smell detection tool PMD \cite{copeland2005pmd}. In the \emph{Write me this code} category, there were 47 conversations and a total of 59 code smell violations in 144 code blocks. Half (50.8\%) of the violations concerned the standard practices of code conventions, a third (37.3\%) related to styles of coding that have an impact on code readability and the remainder (11.9\%) were violations of coding rules that were more likely to lead to errors. 

In the \emph{Improve this code} category, there were 334 conversations. In most cases, the output had fewer total violations and sometimes it was a lot fewer, suggesting that ChatGPT can be used for this purpose. Occasionally the output had more violations typically this was only one or two violations and not of the type that would introduce errors. Most conversations required fewer than 5 attempts; where more were needed it was usually because multiple code snippets were supplied as input.

Moratis et al. observed, however, that their findings were "\emph{inherently optimistic, as it exclusively contains instances of successful interactions with ChatGPT}'' \cite{moratis2024write}. Moreover, it is unclear whether the code quality is better or worse than that of human developers.

Another attempt to apply code smell detection to measure the ability of LLMs to improve the quality of existing code is due to DePalma et al. \cite{depalma2024exploring}, who developed prompts to ask ChatGPT to refactor Java code to improve quality on 8 different quality attributes. Once again, PMD was applied both to the original code and the refactored code.

Liu et al.\cite{liu2024refining} took the idea of code smell measurement one step further to form a self-repairing mechanism. PMD was used once again for Java code but in conjunction with CheckStyle \cite{burn2003checkstyle}. For Python code generation, the tools used were Pylint \cite{thenault2001pylint} and Flake8 \cite{cordasco2010flake8}. 
They classified code quality issues into four categories: (a) Compilation and Runtime Errors, (b) Wrong Output (i.e. functional incorrectness of the generated code), (c) Code Style and Maintainability, and (d) Performance and Efficiency. For each of these categories, they identified the top 10 issues for Java and for Python.  The dataset used was the LMDefect dataset \cite{fan2023automated} of 2033 coding tasks supplemented with coding tasks extracting from LeetCode. The experimental data shows great promise with a repair rate in the range 20\% to 60\%. 
However, fixes can often introduce new quality issues. 

More recently, Feischl and Kern studied the capability of LLMs for translating programs from one language to another by analysing the correctness and quality of the code translated \cite{feischl2025large}. Code smells in the programs before and after translations were detected using smell detection tools PyLint for Python, PMD for Java, Scalastyle for Scala, and Clippy for Rust. In contrast to our metric of \emph{number of violations per solution} (see Section \ref{sec:metrics} for the formal definition), they counted each code smell \emph{at maximum once per code sample}. In other words,  the number of types of smells detected in a sample are used to measure the code quality. Their experiments found that LLMs do not propagate code smells from a source code to the respective translation either directly and semantically. Instead, they appear to resemble code smells which were learned during training for each specific programming language.

In parallel to our work \cite{paul2025does}, Velasco et al. studied the propensity that LLMs have for generating code that smells \cite{velasco2025propense}, but took a completely different approach. They curated a dataset called CodeSmellData from GitHub repositories that contains a set of Python methods such that each method contains at least on code smell as detected by Pylint. These methods contain a total of 142,817 instances of code smells across 13 smell types. Velasco et al. proposed a metric that measures the propensity of generating an instance of code smell $\mu$ as the probability that an LLM generates a specific code $w$ that contains the instance $\mu$ of code smell in a subsequence of tokens in $w$. Given an LLM, the propensity is calculated by analysing the logit from the final layer of an LLM. Using the CodeSmellData dataset, they evaluated two LLMs CodeLlama and Mistral and found that both models are prone to generating 10 out of 13 types of code smells with a propensity (i.e. probability) higher than 0.5.

Xue et al. studied whether removing smells from the training data could improve code quality \cite{xue2025clean}. They tested the presence of code smells in the dataset CodeSearchNet-Python, which is used in training code models. They also evaluated the output of several popular LLMs and found that code smells extensively exist in both the training data and the output from LLMs. They also developed an LLM-based code smell cleaning tool SmellCC to automatically refactor Python code and remove code smells. SmellCC is applied to the dataset CodeSearchNet-Python. Their experiment data shows that SmellCC can remove 91.6\% of code smells while maintain 91.3\% code correctness. The cleaned dataset was used to fine tuning of LLMs DeepSeek-V2 and Qwen-Coder, and reduced code smells in generated code by 79.6\% and 83.1\%, respectively. 

Li et al. \cite{li2026multi} evaluated five coding tools including GitHub Copilot, Amazon CodeWhisperer, Tongyi Lingma, Tabnine, and Codeium on the quality of their code suggestions for five programming languages C, C++, Java, JavaScript, and Python. They assessed the quality of the code generated from 2,613 coding tasks extracted from LeetCode on four quality attributes: correctness, reliability, maintainability, and cyclomatic complexity. Maintainability refers to how easily the code can be understood, modified, and extended. It  is measured by the number of code smells detected by static analysis tools CppCheck and SonarQube. It is worth noting that a subset of the issues detected by static analysis tools were manually validated by two annotators. Their data shows that only 62.7\% of the sampled issues were confirmed by both validators and 11.0\% were false positives. The agreement rate between the two validators was only 91\%, and the Cohen’s kappa coefficient under a binary mapping was only 0.82. This confirms that human agreement on code smell is low.  Moreover, they excluded structural smell issues and naming convention smell issues in the code smell detection. They found that code smells exist in nearly all statically typed language outputs, averaging 1.27 - 1.46 smells per suggestion, and concluded that maintainability issues are pervasive.

\begin{table*}
\caption{Summary of Related Works}
\label{tab:RelatedWorks}
\setlength{\tabcolsep}{2pt}
\centering
\begin{scriptsize}
\resizebox{\linewidth}{!}{
\begin{tabular}{|p{1.0cm}|p{4.0cm}|p{1.3cm}|p{1.0cm}|p{3.2cm}|p{1.6cm}|p{1.2cm}|p{3.8cm}|}
\hline 
\textbf{Work} &\textbf{Aims} &\textbf{Tools} &\textbf{Language} &\textbf{Usage} &\textbf{Dataset (Size)} &\textbf{LLM} &\textbf{Smell Types}\\ 
\hline
\multirow{2}{1.0cm}{Siddiq, et al. 2022} & \multirow{2}{4.0cm}{Investigating the impact of code smells in training datasets on the quality of generated code} & Pylint & \multirow{2}{1.3 cm}{Python} & Detect code smells in training datasets and generated codes & HumanEval (164) & \multirow{2}{1.1 cm}{GPT-Neo, GitHub Copilot} & Implementation smells \\\cline{3-3} \cline{5-5}\cline{8-8}
   & & Bandit & &Detect security smells in training datasets and generated codes & & & Security smells\\\hline
\multirow{2}{1.0 cm}{Moratis, et al. 2024} & Assessing the quality of code generated via iterative conversations in the \emph{Write this code} scenario & \multirow{2}{1.0 cm}{PMD} &\multirow{2}{0.8 cm}{JavaScript} & Detect code smell and measure code quality & DevGPT (47) & \multirow{2}{1.1 cm}{ChatGPT} & \multirow{2}{3.6 cm}{Best practice, code style, error prone} \\ \cline{2-2} \cline{5-6}
   & Assessing the quality improvement of code generated via iterative conversations in the \emph{Improve this code} scenario & & &Comparing code smells before and after refactoring & DevGPT (334) & & \\
\hline
\multirow{1}{1.0 cm}{DePalma, et al. 2024} & Evaluating LLM capability for code refactoring & PMD & Java & Assessing the quality of the code before and after refactoring & Ma et al. \cite{ma2023llms} (40) &ChatGPT & Best practice, code style, design, documentation, error prone, multi-threading, performance, security\\
\hline
\multirow{2}{1.0 cm}{Liu, at el. 2024} & \multirow{2}{4.0 cm}{Evaluating LLM capability for fixing code quality issues} & PMD, CheckStyle & Java & Assessing the quality of generated code & \multirow{2}{0.9 cm}{LMDefect$^+$ (2033)} & ChatGPT &Implementation and design smells\\ \cline{3-4}
   & & Pylint, Flake8 & Python & & & &\\ 
\hline
\multirow{1}{1.0 cm}{Velasco, et al. 2025} &Studying the propsensity (i.e. probability) that LLM-generated code contains code smells &PyLint &Python
&Detecting code smells in Python code when building the CodeSmellData dataset&CodeSmellData (142K) &CodeLlama, Mistral&13 types of method level code smells\\
\hline
\multirow{1}{1.0 cm}{Xue, et al. 2025} &Studying whether cleaning the smell of training data could improve LLM-generated code quality  &SonarQube &Python &Detecting code smells in training dataset and code generated by LLMs. &CodeSearchNet-Python (1000)&CodeLlama, Mistral&All method level code smells (implementation and design smells)\\
\hline 
\multirow{4}{1.0 cm}{Feischl \& Kern 2026} &\multirow{4}{4.0 cm}{Studying LLMs’ capability of code translation considering both correctness and code quality}& PyLint &Python &\multirow{4}{3.2 cm}{Detecting code smells in source language and target language}&Avatar(10) &\multirow{4}{1.2 cm}{GPT-3.5, Codestral, DeepSeek, Llama, etc.}&{Implementation and design smells}\\\cline{3-4}
&&PMD&Java&&CodeNet(9)&&\\\cline{3-4}
&&Scalastyle&Scala&&EvalPlus(7)&&\\\cline{3-4}
&&Clippy&Rust&&EvoEval(4)&& \\
\hline
\multirow{1}{1.0 cm}{Li, et al. 2026} &Assessing the quality of suggested code generated by coding tools&CppCheck, SonarQube
&C, C++, Java, Python, JavaScript &Detecting the number of code smells in generated code as a measure of code maintainability. &LeetCode (2,613)&Copilot, CodeWhisperer, Lingma, Tabnine, Codeium&Excluded structural smells and naming convention smells\\
\hline
\multirow{1}{1.0 cm}{Paul et al. 2025 (This paper)} & Evaluating LLMs on various types of code smells for various types of code to generate and the complexities of coding task & PMD, CheckStyle, DesigniteJava & Java & Assessing the quality of code generated in various scenarios & ScenEval (1000) & Gemini Pro, ChatGPT, Codex, Falcon & Implementation and design smells \\ 
\hline
\end{tabular} 
}
\end{scriptsize}
\end{table*}

Table \ref{tab:RelatedWorks} summarises the related works mentioned in this subsection and contrasts them with the work  reported in this paper. The column \emph{Aims} gives the purpose of the research. The column \emph{Usage} explains how code smell detection techniques achieve that purpose. The columns \emph{Tools}, \emph{Dataset}, \emph{Language}, \emph{LLM} and \emph{Smell Types} give, respectively, the code smell detection tool(s) used, the test dataset used to evaluate the LLM(s) with the size of the dataset in parentheses, the programming language in which the code is generated, the LLM(s) evaluated, and the types of code smell. 

It is worth noting that the eight smells detected by DePalma et al. \cite{depalma2024exploring} are implementation smells. It is not explicitly stated what code smells were detected by Liu et al.'s work. However, we believe that architectural smells were not detected because there is no architectural level code generated by the test cases. For the same reason, architecture code smells were not detected in our work.

\subsection{Research Questions}\label{sec:ResearchQuestions}

The existing works discussed above give some picture of the prevalence of code smells in LLM-generated code, but a question remain: given a quantity of code smells that exist in LLM-generated code, does it mean the code quality is poor and cannot be used? As discussed in Section \ref{sec:Background}, it is difficult to set such a level of standard due to the nature of code smells, which are indicative, subjective and non-functional \cite{beck1999bad}. Therefore, we compare LLM-generated code with the human-written alternative to assess if the quality of LLM-generated code reaches the level of professionally written code. 

Moreover, existing works in the evaluation of ML models on both functional correctness and code quality can be characterised as \emph{benchmarking}. That is, they calculate the average performance of an LLM over a test dataset, which is used as the benchmark. The results heavily depend on the probability distribution of the test data in the benchmark. However, as discussed in Section \ref{sec:ChallengesAndApproach}, developers using an LLM as a coding tool have vastly different distributions of coding problems. It is difficult, if not impossible, to develop a benchmark that reflects the problem distribution for all developers. In fact, there is no code generation benchmark that claims to be representative to the probability distribution for coding problems in the real-world \cite{ghosh2024b}. Our approach to this problem is scenario-based testing and evaluation. 

In particular, we will ask the following open research questions. 

\begin{itemize}
    \item \emph{RQ1.} Does LLM-generated code have a quality comparable to that of human-written code?  
\end{itemize}
    
By human-written code, we mean code written by textbook authors or professional programmers, since we believe that can fairly represent the current best practice. Since our approach, outlined in Section \ref{sec:ChallengesAndApproach}, is to measure code smells, we will ask how does the incidence of smells in LLM-generated code compare with that of human-written code. Since LLMs are mostly used to generate method / function and class level code, this paper focuses on implementation and design code smells while leaving the architectural smells for future work. 

\begin{itemize}
    \item \emph{RQ2.} On which programming topics is LLM-generated code is weaker or stronger in quality compared to human-written code? 
\end{itemize}

Once again, this can be rephrased in terms of code smells. How do code smells of LLM-generated code vary with the question topic? More importantly, on which topics are the smells most worsened or most improved compared to human-written code?

\begin{itemize}          
    \item \emph{RQ3.} How does the quality of LLM-generated code vary with the complexity of the coding task? 
\end{itemize}          

A closely related question would be is the difference in quality compared to human-written code greater for complex coding tasks? Both of these questions can be transformed into corresponding questions on code smells as above. 
    
\begin{itemize}    
    \item \emph{RQ4.} On which quality attributes is LLM-generated code worse compared to human-written code, since that is where research efforts could be directed? 
\end{itemize}

We can rephrase this to ask which code smells are most prevalent in LLM-generated code and whether each smell is more or less common than in human-written code.
   
\begin{itemize}
    \item \emph{RQ5.} How is the correctness of LLM-generated code related to the usability of the code measured by code smell?  
\end{itemize}
    
As discussed in Section \ref{sec:Background}, intensive research in software engineering has clearly demonstrated that code smell is closely related to quality attributes that affect the usability of program code, which include the readability, modifiability, reusability and maintainability (i.e. easiness of evolution of program code). To answer this question, we will separate the codes generated by LLMs according to their correctness, analyse their smells separately, and compare their code smells with the baseline.
To answer this question, we will separate the codes generated by LLMs according to their correctness, analyse their smells separately, and compare their code smells with the baseline. 

\section{Code Smell Detection}\label{sec:Background}

In this section, we will review the notion of code smells as background and explain how they can be detected. We will also explain how the code smell detection tools PMD, Checkstyle and DesigniteJava will be employed in our investigation. 

\subsection{The Notion of Code Smell}

The concept of a \emph{code smell} originated in Fowler's book on refactoring \cite{fowler2018refactoring} having been coined by Beck \cite{beck1999bad}. It was defined as ``\emph{indications that there is trouble that can be solved by a refactoring}'' and ``\emph{certain structures in the code that suggest (sometimes they scream for) the possibility of refactoring}”. The authors described a list of 22 code smells, and how in each case, refactoring methods can help to improve the quality of the program. Since then, the notion of code smell has been intensively studied (see, for example, \cite{santos2018systematic, sobrinho2021systematic, lacerda2020code} for systematic literature reviews) and generalised to software smells \cite{sharma2018survey}.  

Beck and Fowler noted two distinctive aspects of the notion of code smells. Firstly, they are \emph{indicative} rather than ``\emph{precise criteria for flaws in program code}''. The code may not be flawed and may function correctly, but there may be future problems with maintenance, evolution, and reuse\cite{lacerda2020code}.  Secondly, they are \emph{subjective}. It is better to judge smell based on ``\emph{informed human intuition}'' and consequently, ``\emph{human agreement on smell detection is low}'', as has been proven by research \cite{santos2018systematic}. 

The notion of code smell is linked to a number of other software engineering concepts and techniques, as follows:
\begin{itemize}
\item Smells are \emph{indicators} or symptoms of a deeper design problem in the program code, as discussed above.
\item Smells are suboptimal or poor solutions to a coding problem. Bad smells lead to a \emph{technical debt} of needing to find better solutions later. 
\item Smells violate recommended \emph{best practice} for the domain. These include coding conventions and/or software design principles. Therefore, smells can be detected by looking for the violations of best practices. 
\item Smells have a negative impact on the software \emph{quality attributes} that are related to product revision and transition, such as modifiability, readability, testability, reusability, portability, etc. In this way, they make software difficult to evolve, maintain, and reuse, and increase the likelihood of bugs, without themselves being bugs. 
\item As Fowler suggested, smells should and can be eliminated or reduced, for example, by \emph{refactorings}, which are meaning-preserving transformations on the software. 
\item Smells are recurring problems in program code. The patterns of such recurring problems bear similarity to the notion of \emph{anti-patterns}. 
\end{itemize}

\subsection{Types of Code Smells}

Many types of code smells have been defined and investigated in the literature. They can be classified according to a number of different criteria, such as the effect caused, design principles violated, location of the smell, its granularity, etc \cite{sharma2018survey}. In this paper, we adopt the classification proposed by Suryanarayana, Samarthyam and Sharma \cite{suryanarayana2014refactoring, sharma2016configuration} which distinguishes implementation smells from design smells (also known as micro-architectural smells) and architectural smells. 

\subsubsection{Implementation Smells}
    
Implementation smells are concerned with suboptimal implementation choices that make the code unnecessarily complex, difficult to maintain, and harder to understand. We detect and analyse the following: 
\begin{itemize}
  \item \emph{Inconsistent Naming Convention}. Deviations from the recommended naming conventions. 
  \item \emph{Excessive Complexity}. An expression, statement or a method is difficult to understand due to lack of clarity caused by excessive complexity. For example, a statement could be excessively long, an expression could be excessively nested and/or have too many operations, and a method could have too many lines of code, and/or have an excessive list of parameters. 
  \item \emph{Incompleteness}. A piece of code is unfinished with, for example, "TODO" or "FIXME" tags, or a statement is incomplete. For example, a catch block may be missing handling logic, a conditional construct may be missing a terminating \texttt{else} clause, a \texttt{switch} or selector statement may be missing a \texttt{default} case, or more generally, a block of code within curly braces \{\} could contain no executable statements, etc. 
  \item \emph{Redundant Elements}. The presence of duplicate parameters, methods, or code blocks. A method or attribute may have an  unnecessary modifier, such as \texttt{public} where that visibility is already implied, or \texttt{public static final} where \texttt{final} would have been enough.
  \item \emph{Improper Alignment and Placement}. Code is not properly aligned according to coding standards, and/or an entity in the code is misplaced; for example, attributes may not be given in the recommended order. 
  \item \emph{Magic Number}. A numeric literal is used directly in code without being defined as a constant. 
  \item \emph{Dead Code}. Sections of code are no longer executed or provide no value for some other reason.
  \item \emph{Resource Handling}. Inefficiencies in the use of resources. 
  \item \emph{Documentation}. Insufficient comments to explain the code properly. 
\end{itemize}

\subsubsection{Design Smells}
    
Design smells are concerned with design choices, as presented in the program code, that  violate fundamental design principles, such as poor use of object-orientation. They indicate the types of weaknesses that can lead to increased complexity, maintainability issues, and reduced code reusability.
The following are the types of design smells defined by Suryanarayana et al.  \cite{suryanarayana2014refactoring, sharma2016configuration}; these are all detected and analysed in this paper.   

\begin{itemize}
    \item \emph{Abstraction Smell} – Issues related to improper, missing, or unnecessary abstractions, affecting code clarity and reusability.  
    \item \emph{Encapsulation Smell} – Violations of encapsulation principles, such as excessive exposure of internal details or inadequate access restrictions.  
    \item \emph{Modularisation Smell} – Poorly structured modules, including tightly coupled components, improper separation of concerns, and redundant dependencies.  
    \item \emph{Hierarchy Smell} – Problems in class hierarchies, such as deep inheritance trees, improper sub-classing, or lack of adherence to object-oriented principles.  
\end{itemize}

\subsubsection{Architectural Smells}
    
Architectural smells are the weakness in the architectural design
of the system, as presented in the code, that often lead to reduced system flexibility, modification difficulties and maintainability challenges. Typical examples include inappropriate layering and tight coupling between components and subsystems, etc. We will not consider these smells, however, because LLMs have limited capability for generating the entire system architecture and are not normally used for this purpose. 

\subsection{Detecting Code Smells}

Code smell detection has been intensively studied in the software engineering literature; see, for example, \cite{fernandes2016review, haque2018causes, menshawy2021code} for systematic literature reviews. Fowler suggested that detection should be manual based on developer's experience and intuition. However, this is not scalable and repeatable. So automated tools should be used instead. Such tools can be classified into three types: 

\begin{itemize}
  \item \emph{Static Code Analysis Approaches}. 
\end{itemize}  
Tools for static analysis are usually based on either metrics or  pattern-matching. Metrics on program code include Lines of Code (LOC), Number of Attributes per Class (NOA), Number of Methods per Class (NOM), Number of Children Classes (NOC), Depth of Inheritance (DIT), etc. A \textit{metrics-based} tool detects code smells using a combination of these metrics. A drawback of this approach is the arbitrary nature of the threshold values set for the metrics.

A \textit{rule-based} tool, in contrast, defines a set of detection rules based on the syntactic structure of the code. Often these rules are linked to coding conventions and design principles. Sometimes the tool is configurable in that the smells can be specified with editable rules. Usually, they can be seamlessly integrated into existing development workflows. Typical examples of these tools include PMD, Stylecheck, Pylint, DesigniteJava, etc. Recently, such tools have been applied to LLM-generated code; see Section \ref{sec:RelatedWork}. 

\begin{itemize}
  \item \emph{History-based Approaches}. 
\end{itemize} 

The evolution history of the system can be used to analyse the symptoms caused by smells and hence identify the smell. However, only a small number of smells can be detected this way.
  
\begin{itemize}
  \item \emph{ML-based Approaches}. 
\end{itemize}
  
There are two approaches for using machine learning (ML) models \cite{al2020bad,alazba2023deep}. 

The first is to train a model with features based on metrics, like lines of code, cyclomatic complexity, coupling metrics, etc.
This requires large, high-quality labelled datasets. However, these are rare for code smells. Consequently, such ML models have not achieved the performance suitable for practical use \cite{al2020bad,alazba2023deep}. 
For example, the Naive Bayes model reported in \cite{Pecorelli2019Comparing} has low F1-scores for most smells and low precision in particular, indicating a high number of false positives. Moreover, existing ML models for smell detection are binary classification models, i.e. each model only detects one type of smells. 

The second approach is to use LLM models to detect code smells. However, a recent evaluation reported in \cite{Mesbah2025Leveraging} shows low F1 scores for both Llama variants and GPT-4, the latter below 0.04. 

\subsection{Use of Smell Detection Tools}

In this paper, we will use static code analysis tools to detect code  smells in both LLM-generated code and the reference solutions. As with Liu et al. \cite{liu2024refining}, we use PMD \cite{pmd,copeland2005pmd} and Checkstyle \cite{checkstyle,burn2003checkstyle}. They are based on widely recognised coding
conventions: Google Java Style Guide \footnote{https://google.github.io/styleguide/javaguide.html} and the Sun Java Code Convention \footnote{https://www.oracle.com/java/technologies/javase/codeconventions-introduction.html},
respectively. Both of them are capable of detecting and reporting the violations of these rules. However, since both tools are relatively weak in detecting design level code
smells, we also use DesigniteJava \footnote{https://www.designite-tools.com/products-dj}, which detects smells according the design principles violated.

Tables \ref{tab:implementation-smell-rules} and \ref{tab:design-smell-rules} show the smell detection rules provided by each tool used in our work. Columns \emph{Tool Used} and \emph{Detection Rules} give the tool and the smell detection rule used. Readers are referred to the websites of the tools for the definitions of the rules.

Note that none of the tools cover all smells so we need to combine them for maximum coverage. One smell type may be detected by several different rules, even by different tools. The number of violations for such a smell type is calculated by summing up the numbers of violations of different rules. 

Where a rule is implemented by more than one tool, however, the violation is counted only once. We have found in such cases that both tools give the same number of violations for the rule on the same code extract. In Table \ref{tab:implementation-smell-rules}, such cases are indicated by a footnote reference 
\footnote{The result from the tool by applying this smell detection rule is ignored because the same rule is already checked by another tool where the rule may have a different name. Only the result from one tool on the same rule is taken into account.\label{ftn:OnCheckStyle}}. 

\begin{table}[htb]
\caption{Smell Detection Rules Used for Implementation Smells}
\label{tab:implementation-smell-rules}
\setlength{\tabcolsep}{2pt}
\centering
\begin{scriptsize}
\begin{tabular}{|p{1.8cm}|l|l|}
\hline
\textbf{Smell Name} & \textbf{Detection Rule(s)} & \textbf{Tool(s) Used}\\
\hline
\multirow{12}{1.8 cm} {Inconsistent Naming Convention} &{Local Variable Naming Convention} &PMD \\
&{  /Local Variable Name}& { /CheckStyle$^{(\ref{ftn:OnCheckStyle})}$} \\
&Formal Parameter Naming Convention &PMD \\
&Method Naming Convention &PMD\\
&~/Method Name &~/CheckStyle$^{(\ref{ftn:OnCheckStyle})}$ \\
&Class Naming Convention &PMD \\
&GenericsNaming &PMD \\
&AbbreviationAsWordInName &CheckStyle \\
&AbstractClassName &CheckStyle \\
&CatchParameterName &CheckStyle \\
&ConstantName &CheckStyle \\
&IllegalIdentifierName &CheckStyle \\ \hline
\multirow{8}{1.8 cm}{Excessive Complex} &Simplify Boolean Expression &CheckStyle \\
&Simplify Conditional &PMD \\
&Simplify Boolean Return &PMD/CheckStyle \\
&Simplified Ternary &PMD \\
&Line Length &CheckStyle \\
&Method Length &CheckStyle\\
&~/Long Method & ~/DesigniteJava$^{(\ref{ftn:OnCheckStyle})}$ \\
&Excessive Parameter List &PMD/DesigniteJava$^{(\ref{ftn:OnCheckStyle})}$\\ \hline
\multirow{3}{1.8 cm} {Redundancy} &Redundant Import &CheckStyle \\
&Redundant Modifier &CheckStyle \\
&Copy Paste Detector &PMD \\ \hline
\multirow{5}{1.8 cm}{Incompleteness} &Missing Switch Default &CheckStyle \\
&Todo Comment &CheckStyle \\
&Empty Control Statement &PMD \\
&Empty Catch Block &PMD/CheckStyle$^{(\ref{ftn:OnCheckStyle})}$ \\
&EmptyBlock &CheckStyle \\ \hline
\multirow{10}{1.8 cm}{Improper Alignment and placement} &Indentation &CheckStyle \\
&FileTabCharacter &CheckStyle \\
&NeedBraces &CheckStyle \\
&UselessParatheses &PMD \\
&LeftCurly &CheckStyle \\
&RightCurly &CheckStyle \\
&ParenPad &CheckStyle \\
&MethodParamPad &CheckStyle \\
&Variable Declaration Usage Distance &CheckStyle \\
&Declaration Order &CheckStyle \\ \hline
\multirow{1}{1.8 cm}{Magic Number} &Magic Number &CheckStyle \\ \hline
\multirow{5}{1.8 cm}{Dead Code} &Unused Formal Parameter &PMD \\
&Unused Local Variables &PMD/CheckStyle$^{(\ref{ftn:OnCheckStyle})}$\\
&Unused Private Fields &PMD \\
&Unused Private Method &PMD \\
&Unused Imports &CheckStyle \\ \hline
\multirow{2}{1.8 cm}{Resource Handling} &Close Resource &PMD \\
&Avoid Instantiating Objects In Loops &PMD \\ \hline
\multirow{7}{1.8 cm}{Documentation} &Comment Required &PMD \\
&Comment Size &PMD \\
&Comment Content &PMD \\
&Javadoc Method &CheckStyle \\
&Javadoc Type &CheckStyle \\
&Missing Javadoc Package &CheckStyle \\
&Javadoc Variable &CheckStyle \\ \hline
\end{tabular}
\end{scriptsize}
\end{table}

\begin{table}[ht]
\caption{Smell Detection Rules Used for Design Smells }
\label{tab:design-smell-rules}
\setlength{\tabcolsep}{2pt}
\centering
\begin{scriptsize}
\begin{tabular}{|p{20 mm}|l|l|}
\hline
\textbf{Smell Name} & \textbf{Detection Rules} & \textbf{Tool Used} \\
\hline
\multirow{13}{20 mm} {Modularity} &God Class &PMD \\
&Data Class &PMD \\
&Too Many Methods &PMD \\
&Too Many Fields &PMD \\
&Use Utility Class &PMD \\
&Hide Utility Class Constructor &CheckStyle \\
&Broken Modularization &DesigniteJava \\
&Cyclically-dependent Modularization &DesigniteJava \\
&Hub-like Modularization &DesigniteJava \\
&Insufficient Modularization &DesigniteJava \\
&Law of Demeter &PMD \\
&Coupling Between Objects &PMD \\
&Class Fan Out Complexity &CheckStyle \\ \hline
\multirow{7}{20 mm} {Encapsulation} &Visibility Modifier &CheckStyle \\
&Excessive Public Count &PMD \\
&Deficient Encapsulation &DesigniteJava \\
&Final Parameters &CheckStyle \\
&Final Class &CheckStyle \\
&Hidden Field &CheckStyle \\
&Unexploited Encapsulation &DesigniteJava \\ \hline
\multirow{8}{20 mm} {Hierarchy} &Broken Hierarchy &DesigniteJava \\
&Cyclic Hierarchy &DesigniteJava \\
&Deep Hierarchy &DesigniteJava \\
&Missing Hierarchy &DesigniteJava \\
&Multipath Hierarchy &DesigniteJava \\
&Rebellious Hierarchy &DesigniteJava \\
&Wide Hierarchy &DesigniteJava \\
&Dependency Cycles btw Packages &DesigniteJava \\ \hline
\multirow{4}{20 mm}{Abstraction} &Imperative Abstraction &DesigniteJava \\
&Multifaceted Abstraction &DesigniteJava \\
&Unnecessary Abstraction &DesigniteJava \\
&Unutilized Abstraction &DesigniteJava \\
\hline
\end{tabular}
\end{scriptsize}
\end{table}
Since empirical studies have found it difficult to set a limit on the number of violations for the code still to be of good quality, we will count the number and compare it with a baseline; this reflects current practice. 

\section{Test System for Code Smell Analysis and Evaluation}\label{sec:TestSystem}

Our experiments with LLMs need to be automated and we do this by applying the methodology of datamorphic testing proposed by Zhu et al. \cite{zhu2019a}, \cite{zhu2020}. This treats software testing as a systems engineering problem and it encourages both efficient management of test resources and the evolution of the test facilities alongside that of the software under test. 

According to the methodology, a test system comprises two types of artefacts: \emph{test entities} and \emph{test morphisms}. The former are objects and documents involved in testing, such as test data, test datasets, test results, etc. while the latter are operations that manipulate and/or generate these entities to perform testing tasks. This methodology is supported by the test automation environment Morphy \cite{zhu2020}.

Morphy provides a Java framework in which a test system can be implemented as a Java class (more precisely, a hierarchy of Java classes) that consists of a set of attributes representing the test entities and a set of methods representing test morphisms. They are both annotated with metadata so that they can be recognised  by Morphy, seamlessly integrated with Morphy's testing tools, and applied to achieve test automation. The following types of test morphisms are recognised by Morphy.

\begin{itemize}
    \item \emph{Seed Maker}: Generates initial test cases from other entities.
    \item \emph{Datamorphism}: Transforms existing test cases into new ones.
    \item \emph{Metamorphism}: Verifies the correctness of test cases and returns a Boolean result.
    \item \emph{Test Set Filter}: Adds or removes test cases from a test set.
    \item \emph{Test Set Metric}: Maps a test set to a real value, such as test adequacy.
    \item \emph{Test Case Filter}: Maps a test case to a Boolean value. It can be used to determine whether the test case should be retained in the test set.
    \item \emph{Test Case Metric}: Assigns a real-valued metric to individual test cases (e.g., complexity).
    \item \emph{Analyser}: Examines the test set and produces a test report.
    \item \emph{Executer}: Runs the program under test using inputs from test cases and captures the outputs.
\end{itemize}

Given a test system implemented in Java, Morphy supports test automation at the following three levels. 

\begin{itemize}
    \item \emph{Action}: Executes a single test activity using test morphisms, built-in functions or tools. 
    \item \emph{Strategy}: Applies test strategies, which are algorithms with test morphisms and test entities as parameters. 
    \item \emph{Process}: Runs test scripts of a high-level of abstraction. Such scripts can be obtained by recording interactive operations of Morphy, which can be edited, or even manually written, and replayed. 
\end{itemize}

Our test system extends that for a previous experiment, to analyse correctness and completeness of LLM-generated code \cite{ghosh2024} and uses the same benchmark, ScenEval. We define, however, a new test entity \emph{code smell report} and the following new test morphisms. 

\begin{itemize}
\item \emph{Test Executors: LLM invokers}. 
Four test executors for each of the four LLM models \emph{Gemini Pro}, \emph{Codex}, \emph{Falcon7B}, and \emph{ChatGPT}. The last of these has been inherited from the previous work \cite{ghosh2024} but the first three are new. Each executor submits the query to its respective LLM via an API call, and then extracts the Java code from the response text and saves it to a file for further analysis.

\item \emph{Analysers: Code Smell Detector Invokers}. 
Three new test morphisms: \emph{PMD-analyser}, \emph{Checkstyle-analyser} and \emph{DesigniteJava-Analyser}, which invoke the corresponding static code analysis tools PMD, Checkstyle and DesigniteJava, and save the code smell reports into files.

\item \emph{Test Set Metrics: Code Smell Statistical Analysers}.  
Three analysers: \emph{Violations per Solution}, \emph{Baseline Violations per Solution} and \emph{Increase Rate to Baseline}, which perform statistical analysis on the code smell report files.
\end{itemize}

\begin{figure}[htbp]
\begin{center}
\includegraphics[width=\columnwidth]{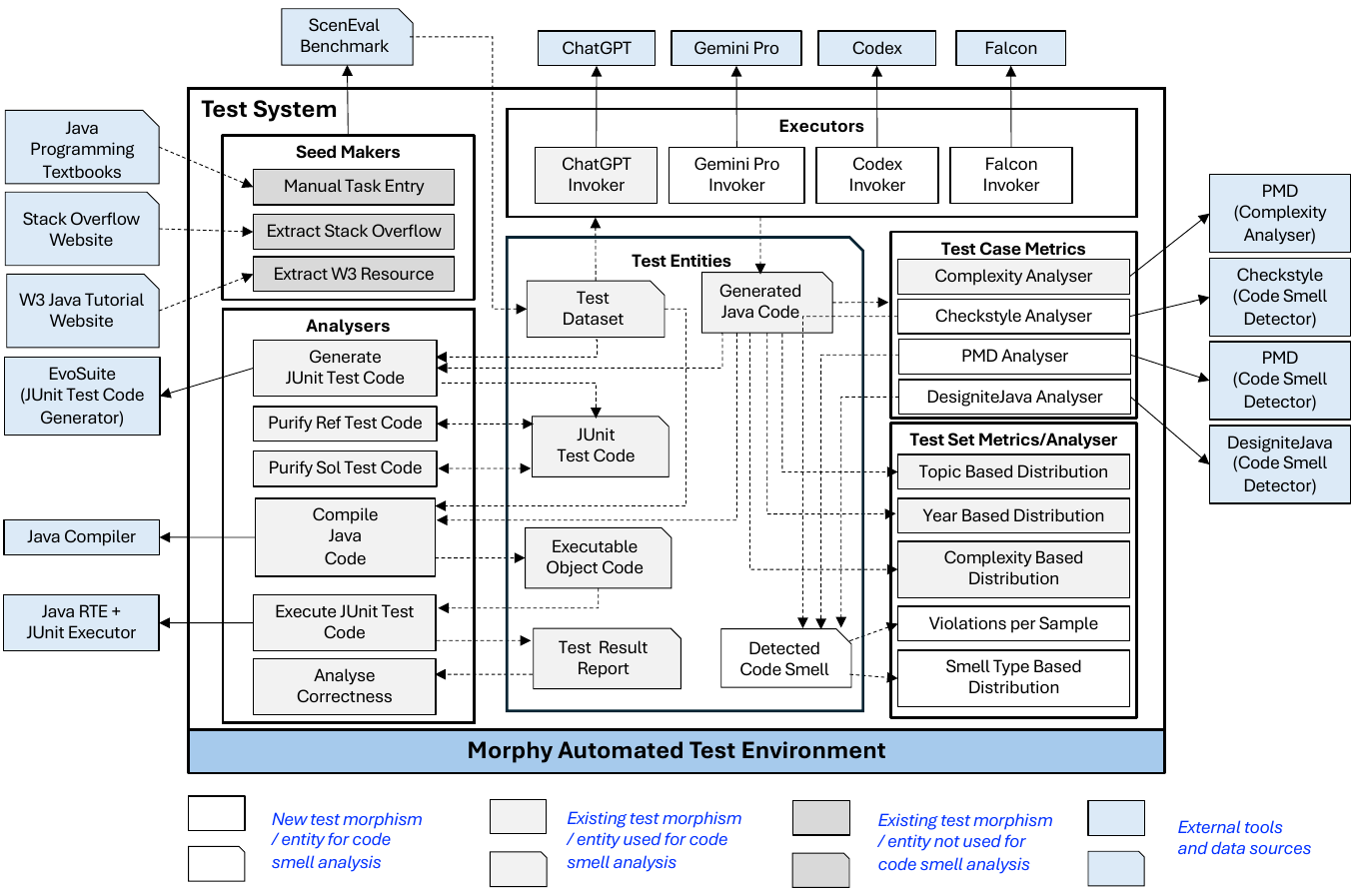}
\caption{Structure of The Test System}
\label{fig:TestSystemStructure}
\end{center}
\end{figure}


Fig. \ref{fig:TestSystemStructure} shows the structure of the test system. Test morphisms and entities inherited from, and explained in, our previous work \cite{ghosh2024} are shown in   grey. New test morphisms and entities are shown in white. External tools invoked by them are shown in blue. Some of these are implemented in Python but invoked through a simple Python2Java interface.

\section{Design of the Experiments}\label{sec:ExperimentDesign}

Before discussing the experiment process, we will first review the LLMs, the benchmark and the platform. 

\subsection{Subject LLMs}

The LLMs ChatGPT/GPT-3.5, Gemini Pro 1.0, Falcon-7B, and Codex were chosen as the subject LLMs based on their availability and prominence at the time when the experiments were initiated in 2023. Our objective was to evaluate a representative mix of widely used proprietary models (ChatGPT, Gemini Pro, and Codex) and an open-source alternative (Falcon-7B). These models were among the most accessible and influential code-generation systems available during the study design phase.

To ensure a fair comparison, all evaluated LLMs were prompted using the exact same task descriptions from the ScenEval benchmark. The prompt supplied to each model consisted strictly of the programming task description, without any few-shot examples. To evaluate the out-of-the-box capability of each model, we maintained the default decoding parameters (such as temperature and $top\_p$) provided by the respective model platforms at the time of experimentation, without manual tuning.  Table \ref{tab:LLMs} presents basic information about these four LLMs we have studied. 

\begin{table}[htb]
\caption{Large Language Models Evaluated}
\label{tab:LLMs}
\centering
\scriptsize
\begin{tabular}{|l|l|l|l|}
\hline
\textbf{Name} & \textbf{Year} & \textbf{Version} & \textbf{Size} 
\\ \hline
Gemini Pro & 2023 & Gemini Pro 1.0 & Unknown \\ \hline
Falcon & 2023 & Falcon-7B & 7B \\ \hline
ChatGPT & 2023 & GPT-3.5-turbo & Unknown \\ \hline
Codex & 2021 & GPT-3 (Codex) & 12B \\ \hline
\end{tabular}
\end{table}

\subsection{Benchmark} 

The benchmark ScenEval \cite{ghosh2024} contains more than 12,000 test cases of Java programming tasks. These test
cases were curated from textbooks, online tutorial websites and the professional programming knowledge-sharing website Stack Overflow. In contrast to other benchmarks for code generation (see e.g. \cite{ghosh2024b}), ScenEval has two distinctive features which make it ideal for our purpose. 

\begin{enumerate}
    \item Each test case is accompanied by the Java code for a reference solution, typically textbook answers and highly-rated Stack Overflow posts. As discussed in \ref{sec:ChallengesAndApproach}, their code quality represents the state of current practice so they enable us to establish a baseline of code smell for human-written Java code.
    \item Each test case is also accompanied by metadata that specifies topic, complexity, source of the task, etc. This enables us to analyse the relationship between these concepts and code smell so that we can answer the research questions.
\end{enumerate}

Our test dataset, sampled at random from the ScenEval benchmark, contains equal quantities (500 each) from textbooks and Stack Overflow. The other statistics are given in Table \ref{tab:StatisticsOfTestDataset}; Input Length and Complexity denote the number of words in the task description and the cyclomatic complexity of the reference solution.

\begin{table}[htb]
\caption{Statistical Characteristics of the Test Dataset}
\label{tab:StatisticsOfTestDataset}
\centering
\begin{scriptsize}
\begin{tabular}{|l|c|c|}
\hline
\textbf{Feature} & \textbf{\#Textbook Tasks} & \textbf{\#Real Tasks}\\ \hline\hline
\# Topics & 25 & 18\\ \hline
\# Tasks per Topic (average) & 20.00 & 27.78\\ \hline
\# Tasks per Topic (max) & 79 & 61\\ \hline
\# Tasks per Topic (min) & 8  & 5\\ \hline\hline
Input Length (average) & 18.55 & 21.54 \\\hline
Input Length  (max) & 35 & 31 \\ \hline
Input Length  (min) & 11 & 16\\ \hline\hline
Complexity (average) & 3.448 &3.200\\ \hline
Complexity (max) & 6 & 5\\ \hline
Complexity (min) & 1 &1\\ \hline\hline
Number of Coding Tasks & 500 &500\\ \hline\hline
\end{tabular}
\end{scriptsize}
\end{table}

For each benchmark task, exactly one solution was generated per model. Consequently, this study utilizes a strict pass@1 evaluation metric. This setup reflects a realistic developer workflow, simulating a scenario where a user requests code from an LLM and directly evaluates the first response returned.

\subsection{Experiment Platform} 

\begin{figure}[htb]
\centering
\includegraphics[width=8cm]{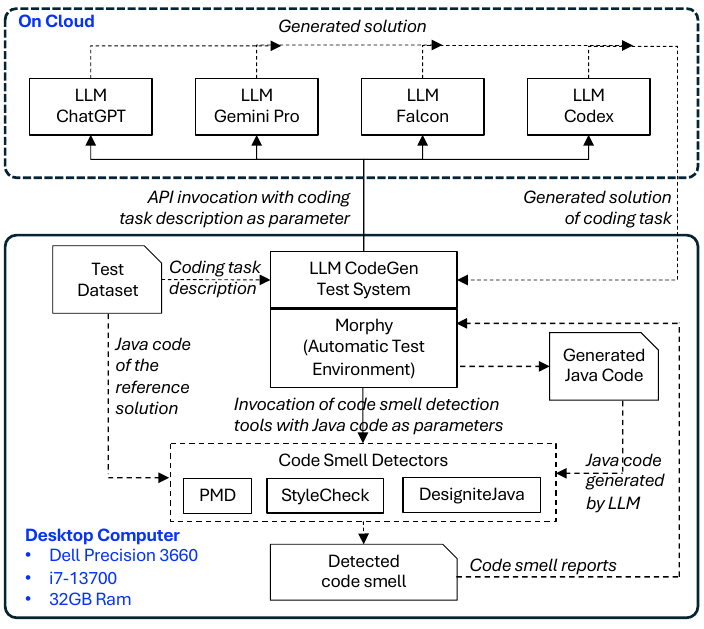} 
\caption{The Experiment Setup}
\label{fig:experiment-setup}
\end{figure}

The experiment was conducted with the automated test environment Morphy
\cite{zhu2019a, zhu2020} running on a desktop computer and is illustrated in Fig.~\ref{fig:experiment-setup}. The LLM models under test were invoked through API calls as discussed in Section \ref{sec:TestSystem} and the smells of the LLM-generated code were analysed with PMD, Checkstyle and DesigniteJava as discussed in Section \ref{sec:Background}.

\subsection{Experiment Process} 

The process of the experiment consists of the following steps. 

\begin{enumerate}
  \item \emph{Constructing Test Dataset}: A test dataset containing 1000 coding tasks was constructed by random sampling of the ScenEval benchmark. 
  \item \emph{Analysing Smells of Reference Solutions}: For each coding task in the test dataset, the Java code for the reference solution is analysed by invoking the code smell detection tools PMD, Checkstyle and DesigniteJava. The reports on the violations of smell detection 
  rules are saved into separate files for further statistical analysis to establish the baseline. 
  \item \emph{Invoking LLMs}: For each coding task in the test dataset, the LLMs under test are queried through API invocations and the solutions returned were collected. The Java code in the returned text is separated from the explanation text and saved into a .java file. 
  \item \emph{Analysing Smells of LLM-generated Code}: The Java codes generated by the LLMs are analysed via invocations of the code smell detection tools PMD, Checkstyle and DesigniteJava. The violations of the detection rules are saved into code smell report files. 
  \item \emph{Analysing Correctness of Generated Code}: For each LLM-generated Java code, the correctness of the code is determined by running test data on both the LLM-generated code and the reference solution; see \cite{ghosh2024} for details. 
  \item \emph{Analysing Complexity of Generated Code}: The complexity of the LLM-generated code is measured the same way as in our previous work. \cite{ghosh2024}.
  \item \emph{Statistical Analysis}: The code smell report files are parsed, and statistical analysis is performed on various subsets of the test dataset to answer the research questions. 
  Each subset represents a different scenario in the use of LLMs. 
  The smell detection rules are partitioned into subsets according to their type, where needed to answer a research question. 
  
 Given a subset  $T$ of the coding tasks and a subset $S$ of smell detection rules, the following statistical data are calculated. 
  \begin{enumerate}
  \item The number of violations of detection rules per 
  solution, denoted by $VS^\mathcal{M}_S(T)$, is calculated for each LLM $\mathcal{M}$ using Equ. (\ref{equ:VS}) . 
  \item The baseline for the test subset $T$ w.r.t. smell detection rules in $S$, i.e. the number of violations of smell detection rules per solution, denoted by $VS^\mathcal{B}_S(T)$, is calculated from the code smell reports of reference solutions using Equ. (\ref{equ:VS_Baseline}). 
  \item The increase rate of smells for LLM-generated code with respect to the baseline, denoted by $Inv^\mathcal{M}_S(T)$, is calculated from $VS^\mathcal{M}_S(T)$ and  $VS^\mathcal{B}_S(T)$ using  Equ. (\ref{equ:VS_Increase}). 
   \end{enumerate}
\end{enumerate}

These three equations are implemented as test set metrics and formally defined in the next subsection. 

\subsection{Metrics of Performance}\label{sec:metrics}

It is worth noting that a piece of program code may violate a code smell rule multiple times at different locations in the code. Moreover, at the same location in a program code, it is possible that multiple code smell rules are violated. To measure the strength of code smells for a piece of code, such multiple violations of same rule at different locations and violations of different rules at the same location are considered as different violations. This is the common practice in the research on code smells. This is also used in this paper. 

Let $s$ be any given smell detection rule and $c$ be a given Java code sample. Formally, we write $V_{s}(c)$ to denote the set of all violations of the rule $s$ detected in the Java code $c$ at all locations in $c$. Therefore, $|V_{s}(c)|$ is the number of violations of coding rule $s$, indicating how strong the code program code $c$ smells according to rule $s$. The larger the value of $|V_{s}(c)|$, the stronger the code smell, and indicating that $c$ is of a poorer quality in terms of the violations of coding rule $s$. 

Now, we write $V_{S}(c)$ to generalise the notation $V_{s}(c)$ to a set of smell detection rules $S \neq \emptyset$. That is, 
\[V_{S}(c)=\bigcup_{s \in S}V_{s}(c).\] 
Informally, it is the set of $c$'s \emph{violations of all code smell rules} $s \in S$. Thus, $|V_{S}(c)|$  is the sum of $|V_{s}(c)|$ over all smell detection rules $s$ in $S$. Thus, we have that 

\[|V_{S}(c)| = \sum_{s \in S}|V_{s}(c)|.\] 

Given a sample set of program codes $P \neq \emptyset$, the metric \emph{smell violations per solution}, denoted by $VS_S(P)$, can be formally defined as follows. 

\begin{Definition}(Smell Violations Per Solution (VS))\label{def:VS}

Let $S\neq \emptyset$ be a set of code smell rules, $P \neq \emptyset$ be a sample set of program codes. The \emph{smell violations per solution} of $P$ w.r.t. $S$, denoted by $VS_S(P)$, is the average $|V_{S}(c)|$ over the set of program code $c \in P$, which is formally defined by the following equation. 

\begin{equation}\label{equ:VS}
VS_{S}(P)=\frac{\sum_{c \in P}{|V_{S}(c)|}}{|P|}
\end{equation}
\qed
\end{Definition}

Informally, $VS_{S}(P)$ measures the average strength of code smell over the sample set $P$ of program code w.r.t. coding rules $S$. Given a set $S$, the larger the value of $VS_{S}(P)$, the stronger the code smell of the programs in $P$ on average, and thus, the poorer quality on average for the programs in $P$. 

In the sequel, we are interested in the sample programs that are generated by an LLM $\mathcal{M}$ from a set $T$ of coding tasks and we compare them with a baseline sample programs provided by the benchmark as the reference solutions for the same coding tasks. Let $t$ be a given coding task. We write $\mathcal{M}(t)$ to denote the program code generated by an LLM model $\mathcal{M}$ on coding task $t$ and $\mathcal{R}(t)$ to denote its reference solution provided by the benchmark for coding task $t$. Let $T \neq \emptyset$ be a set of coding tasks, we introduce the following notations. 

\[VS_{S}^\mathcal{M}(T)=VS_{S}(\{\mathcal{M}(t) | t \in T \})\]
\[VS_{S}^\mathcal{B}(T)=VS_{S}(\{\mathcal{R}(t) | t \in T\})\]

In the above equations, the set $T$ of coding tasks can be a subset of the benchmark, such as those for a specific coding topic or complexity of the coding tasks in the test dataset. Informally, $VS_{S}^\mathcal{M}(T)$ and $VS_{S}^\mathcal{B}(T)$ are the average code quality of the code generated by the LLM $\mathcal{M}$ from the test dataset $T$ and the reference solutions provided by the benchmark for coding tasks in $T$, respectively. By definition, they can be also calculated by the following equations. 

\begin{equation}\label{equ:VS_LLM}
VS_{S}^\mathcal{M}(T)=\frac{\sum_{t \in T}{\sum_{s \in S}{\|V_{s}(\mathcal{M}(t))\|}}}{\|T\|}
\end{equation}

\begin{equation}\label{equ:VS_Baseline}
VS_{S}^\mathcal{B}(T)=\frac{\sum_{t \in T}{\sum_{s \in S}{\|V_{s}(\mathcal{R}(t))\|}}}{\|T\|}
\end{equation}

%
%
%
%
%
%
%

Since the number of smell violations per solution indicates the average strength of code smell, the higher values mean poorer code quality on average. To compare the quality of LLM-generated code against a baseline, we will also measure the \emph{increase rate} of code smells, as defined below. 

\begin{Definition}(Increase Rate of Code Smells)

The \emph{increase rate} of code smells for LLM model $\mathcal{M}$ with respect to the baseline on a set $S$ of smell detection rules over a set $T$ of coding tasks is denoted by $Inc_{S}^\mathcal{M}(T)$, which is formally defined by the following equation. 

\begin{equation}\label{equ:VS_Increase}
Inc_{S}^\mathcal{M}(T)=\frac{VS_{S}^\mathcal{M}(T)-VS_{S}^\mathcal{B}(T)}{VS_{S}^\mathcal{B}(T)}
\end{equation}
\qed
\end{Definition}

Positive values for this quantity mean that LLM-generated code is of lower quality than the reference solution. Negative values means the quality of LLM-generated code is better than the reference solutions. 

\section{The Results}\label{sec:Results}

In this section, we report the data collected from our experiments and answer each of the research questions with a statistical analysis of the data.  

\subsection{RQ1: Prevalence of Code Smells}

\begin{figure}[htb]
\centering
\includegraphics[width=8cm]{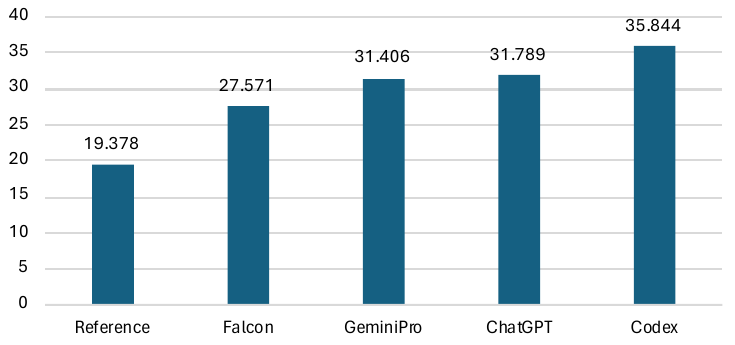}\\
\scriptsize{(a) Smell Violations per Solution}\\
\vspace{0.2cm}
\includegraphics[width=8cm]{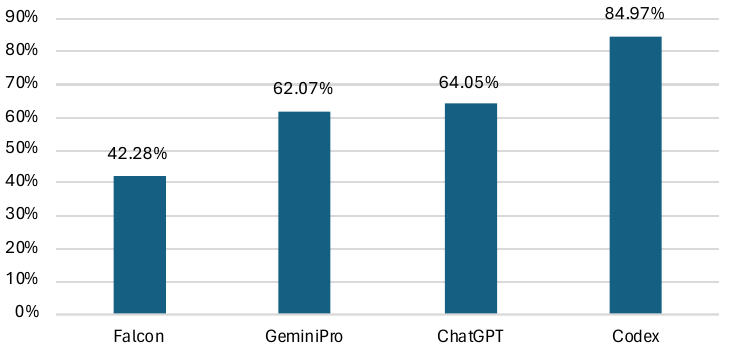}\\
\scriptsize{(b) Increase Rate (\%) of Code Smells}
\caption{All Code Smells Detected on the Whole Test Dataset}
\label{fig:TotalViolations}
\end{figure}

Research question \emph{RQ1} is concerned with the overall
quality of the code generated by the LLMs. To answer this question, we calculated the VS on all smell detection rules over the whole test dataset for each LLM and compared it with the baseline. The results are given in Table \ref{tab:OverallVS} and visualised in Fig. \ref{fig:TotalViolations}. 

\begin{table}[h]
\caption{Overall Code Smells in LLM-generated Code}
\label{tab:OverallVS}
\resizebox{\linewidth}{!}{
\begin{tabular}{|l|c|c|c|c|}
\hline
\textbf{LLM Model} &\textbf{Avg VS} &\textbf{Inc.Rate (\%)} &\textbf{MoE(\%)} &\textbf{p-value}\\
\hline
Reference &19.378 &N/A &6.045 &N/A \\\hline
Falcon &27.571 &42.28\% &4.654 &1.0749E-19\\\hline
Gemini Pro &31.406 &62.07\% &4.284 &6.0923E-38\\\hline
ChatGPT &31.789 &64.05\% &4.029 &3.0557E-42\\\hline
Codex &35.844 &84.97\% &3.708 &3.9380E-68\\\hline
\end{tabular}
}
\end{table}

As shown in Fig. \ref{fig:TotalViolations}(a), all four LLMs have stronger code smells than the baseline with Falcon performing the best (VS=27.571) and Codex the worst (VS=35.844). Moreover, the increase rate of code smells varies significantly in the range from 42.28\% for Falcon to 84.97\% for Codex. 

The data are of high confidence as shown in column \emph{MoE (\%)} of Table \ref{tab:OverallVS} for the percentage margins of errors in the statistical data at confidence level 95\%.  

Therefore, we have the following observations. 

\begin{Observation} LLM-generated code is of poorer quality in terms of the smell violations per solution when compared to the human-written code. 
\end{Observation}

\begin{Observation} 
The rate of increase in code smells of LLM-generated code when compared to the human-written code varies significantly with the choice of LLM. 
\end{Observation}

Table \ref{tab:OverallVS} also shows, in column \emph{p-value}, the results of t-tests with the null hypothesis below: 

\emph{$H_0$: There is no significant difference between human-written  code and LLM-generated code on code smells}. 

Since the p-values are in the range between 3.93804E-68 and 3.05572E-42, we can confidently reject the null hypothesis. Thus, the conclusion of Observation 1 is of high statistical inference strength. 

\subsection{RQ2: Variation of Code Smells by Topic} 

Research question \emph{RQ2} aims to identify the types of tasks for which LLM-generated code is of poor quality so that improvement can be directed to these tasks. To answer this
question, we divide up the test dataset according to the topic of the code to be generated and then calculate the VS on subsets of these coding tasks with all smell detection rules. 

Table \ref{tab:violations_comparison} presents the VS values for various programming
topics and the increase rates for each LLM model. 

\begin{table*}[ht]
    \caption{Code Smell by Code Topics} 
    \label{tab:violations_comparison}
    \centering
    \includegraphics[width=18.5cm]{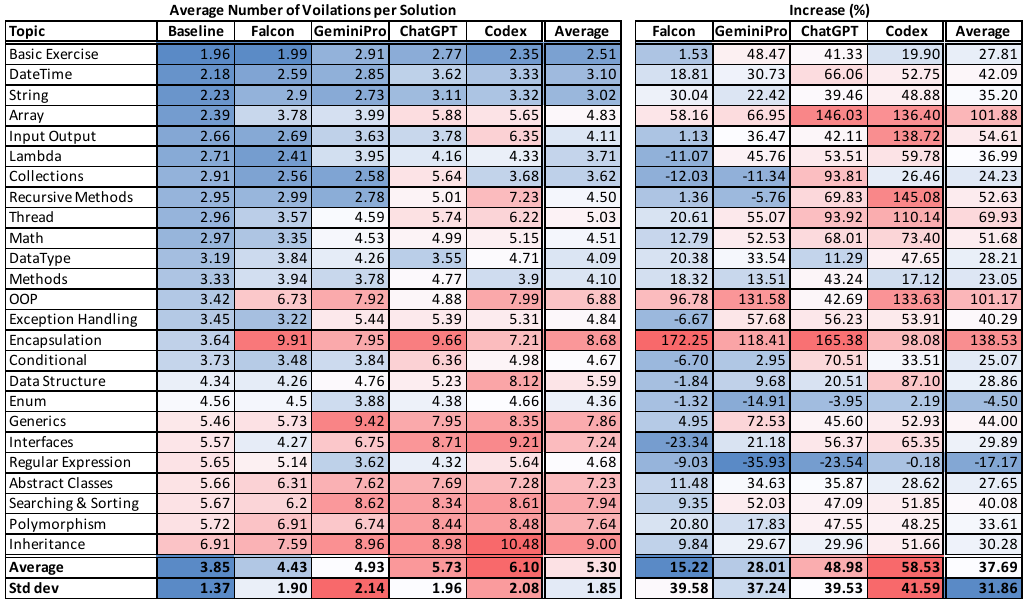}
\end{table*}

From the experiment data, the following observations can be made. 

First, the experiment data shows that LLMs performed differently on the program topics. Although the strength of code smells for the baseline also vary over code topics, the LLMs have a higher standard deviation on VS values when compared to the baseline (1.37). The standard deviations for LLMs are 1.90 for Falcon, 1.96 for ChatGPT, 2.08 for Codex and 2.14 for Gemini Pro, respectively. On average over all LLMs, the standard deviation is 1.85, which is an increase of 31.86\% compared to the baseline. This indicates that LLMs have a robustness problem in maintaining consistency in code quality. In this respect, Falcon is the best (1.90) and
Gemini Pro the worst (2.14).  

\begin{Observation} 
Over different coding topics, the degree of code smells in LLM-generated code varies significantly more than human-written  code. 
\end{Observation}

Second, not only does the degree of code smells vary over different code topics, there is a pattern to the variation. We analysed the Pearson correlation coefficients between the baseline VS values and those of LLM-generated code over various code topics. We found that there is a strong correlation between them. The Pearson correlation coefficients are 0.6652 for Falcon, 0.7249 for Gemini Pro, 0.7188 for ChatGPT and 0.7664 for Codex, respectively, and 0.7876 on average over all LLMs. Therefore, we have the following observation. 

\begin{Observation}
The coding topics with strong code smells in human-written code are the same as the coding topics with strong code smells in LLM-written code. 
\end{Observation}

However, we found no strong correlation between the baseline VS values and the LLM's increase rates of code smells. In fact, the Pearson correlation coefficients are negative, between -0.1592 and -0.3808. 

\begin{table*}[htb]
    \caption{The Best/Worst Topics And The Most Improved/Worsened Topics}
    \label{tab:BestWorstTopics}
    \centering
    \begin{scriptsize}
    \resizebox{\linewidth}{!}{
    \begin{tabular}{|l|l|l|l|l|}
\hline
\textbf{Model} &\textbf{Best Topics} (VS) &\textbf{Worst Topics} (VS) &\textbf{Most Improved Toipics} (Inc \%) &\textbf{Most Worsened Topics} (Inc \%) \\ \hline
\multirow{3}{*}{Baseline} & Basic Exercise (1.96) &Searching \& Sorting (5.67) &N/A &N/A \\
 &DateTime (2.18) &Polymorphism (5.72) &N/A &N/A \\
 &String (2.23) &Inheritance (6.91) &N/A &N/A \\ \hline
\multirow{3}{*}{Falcon} &Basic Exercise (1.99) &Polymorphism (6.91) &Interfaces (-23.34) &Array (58.16) \\
 &Lambda (2.41) &Inheritance (7.59) &Collections (-12.03) &OOP (96.78) \\
 &Collections (2.56) &Encapsulation (9.91) &Lambda (-11.07) &Encapsulation (172.25) \\ \hline
\multirow{3}{*}{Gemini Pro} &Collections (2.58) &Searching \& Sorting (8.62) &Regular Expression (-35.93) &Generics (72.53) \\
 &String (2.73) &Inheritance (8.96) &Enum (-14.91) &Encapsulation (118.41) \\
 &Recursive Methods (2.78) &Generics (9.42) &Collections (-11.34) &OOP (131.58) \\ \hline
\multirow{3}{*}{ChatGPT} &Basic Exercise (2.77) &Interfaces (8.71) &Regular Expression (-23.54) &Thread (93.92) \\
 &String (3.11) &Inheritance (8.98) &Enum (-3.95) &Array (146.03) \\
 &DataType (3.55) &Encapsulation (9.66) &DataType (11.29) &Encapsulation (165.38) \\ \hline
\multirow{3}{*}{Codex} &Basic Exercise (2.35) &Searching \& Sorting (8.61) &Regular Expression (-0.18) &Array (136.40) \\
 &String (3.32) &Interfaces (9.21) &Enum (2.19) &Input Output (138.72) \\
 &DateTime (3.33) &Inheritance (10.48) &Methods (17.12) &Recursive Methods (145.08) \\ \hline\hline
\multirow{3}{*}{\textbf{Average}} &Basic Exercise (2.51) &Searching \& Sorting (7.94) &Regular Expression (-17.17) &OOP (101.17) \\
 &String (3.02) &Encapsulation (8.68) &Enum (-4.50) &Array (101.88) \\
 &DateTime (3.10) &Inheritance (9.00) &Methods (23.05) &Encapsulation (138.53) \\
	\hline
	\end{tabular}
	}
	\end{scriptsize}
\end{table*}

Finally, the data also shows that LLMs are more likely to worsen the code smells on more advanced coding topics. Table \ref{tab:BestWorstTopics} shows the best and worst three topics as well as the most improved and worsened three topics by each LLM and on average over all studied LLMs. The best topics (i.e. the strength of code smells decreased) are \emph{Basic Exercise},  \emph{String}, \emph{DateTime}; while the worst topics (i.e. the strength of code smells increased) are \emph{Searching and Sorting}, \emph{Encapsulation}, \emph{Inheritance}, \emph{Polymorphism}, \emph{Interfaces}, and \emph{Generics}. 

There are a few topics on which LLMs improved the code quality in terms of code smells. These include \emph{Regular Expression} by all LLMs and \emph{Enum} improved by Gemini Pro, ChatGPT and Codex, \emph{Collections} by Falcon and Gemini Pro, \emph{Interfaces} and \emph{Lambda} by Falcon and \emph{Methods} by Codex, and \emph{DataType} by ChatGPT. The highest improvement is 35.93\% by Gemini Pro on the topic of \emph{Regular Expressions}. 

On average over all LLMs studied, the most worsened topics are \emph{Encapsulation} by 138.53\%, \emph{Array} by 101.88\%, and \emph{OOP} by 101.88\%. The largest increase rate of code smell is 165.38\% by ChatGPT on the topic of \emph{Encapsulation}. Thus, we have the following observations. 

\begin{Observation}
The LLM-generated code has the best quality on basic coding topics, and the worst on advanced coding topics.
\end{Observation}

\begin{Observation} 
The LLM-generated code can have better quality in comparison with human-written  code on certain coding topics, while it can also have significantly worse code quality, especially on advanced coding topics. 
\end{Observation}

\subsection{RQ3: Variation of Code Smells by Complexity} 

\begin{figure}[htb]
\centering
\includegraphics[width=7cm]{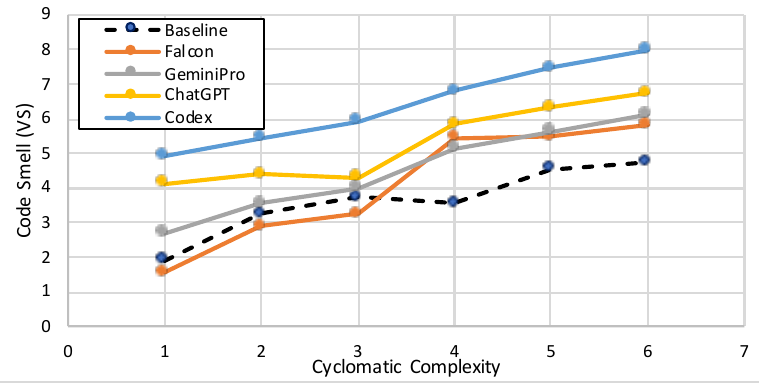}\\
\includegraphics[width=7cm]{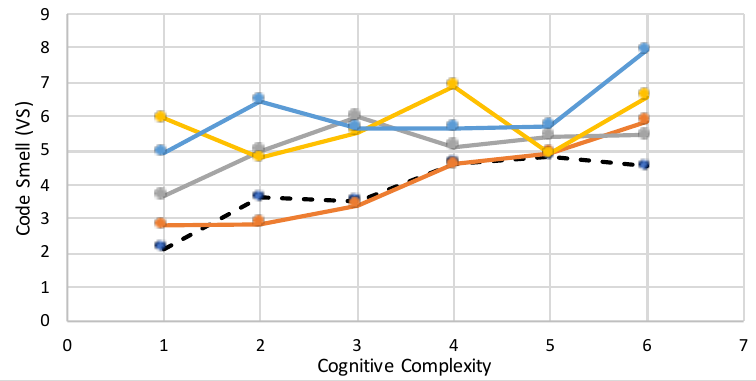}\\
\includegraphics[width=7cm]{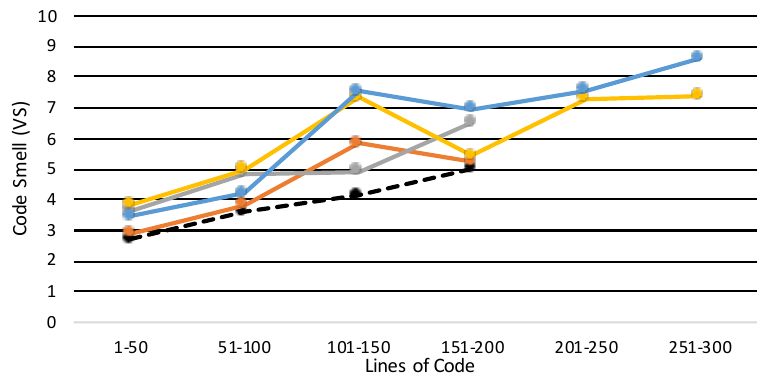}
\caption{Variation of Code Smells by Complexity}
\label{fig:VariationOfSmellWithComplexity}
\end{figure}

Research question \emph{RQ3} is concerned with how code quality
varies with the complexity of the coding tasks. To answer this
question, we calculated the VS on subsets of test cases that were
formed according to the complexity of the coding tasks. Here, the complexity of a coding task is measured on the complexity of the reference solution provided by the benchmark ScenEval. Three different complexity metrics were used: cyclomatic complexity, cognitive complexity and lines of code. The results are presented in Fig. \ref{fig:VariationOfSmellWithComplexity} 
graphically. 

From Fig. \ref{fig:VariationOfSmellWithComplexity},  it can be seen clearly that the VS tends to increase with each of three different metrics of complexity. This is confirmed by the Pearson Correlation coefficients between VS and complexity; see Table \ref{tab:ComplexityCorrelation}. For cyclomatic complexity, the Pearson correlation coefficients are all strongly positive (above 0.9) for each LLM as well as the baseline. For lines of code, the
coefficients are a bit lower but still very high (in the range between 0.8685 and 0.9952 except ChatGPT (0.6347). For cognitive complexity, the results are mixed: very strong for Falcon (0.9754), but much lower for ChatGPT (0.3025), and around 0.6 for Gemini Pro and Codex, and 0.6894 on average over all LLMs. 

\begin{table}[htb]
    \caption{Correlations Bwt VS and Coding Task Complexities} 
    \label{tab:ComplexityCorrelation}
    \centering
    \begin{scriptsize}
    \begin{tabular}{|@{ }l@{ }|@{ }c@{ }|@{ }c@{ }|@{ }c@{ }|@{ }c@{ }|@{ }c @{ }|@{ }c@{ }|}
        \hline
        \textbf{ Complexity} & \textbf{ Baseline} & \textbf{ Falcon} & \textbf{ GeminiPro} & \textbf{ ChatGPT} & \textbf{ Codex } &\textbf{ Average}\\
        \hline
{ Cyclomatic} &0.9344 &0.9555 &0.9916 &0.9485 &0.9962 &0.9653 \\ \hline
{ Cognitive} &0.8800 &0.9754 &0.6312 &0.3025 &0.6578 &0.6894 \\ \hline
{ Lines of Code} &0.9952 &0.8694 &0.9532 &0.6347 &0.8900 &0.8685 \\
        \hline
        \end{tabular}
        \end{scriptsize}
\end{table}

Therefore, we have the following observations. 

\begin{Observation} 
For both human-written  code and LLM-generated code, the strength of code smells increases with the complexity of coding tasks. 
\end{Observation}

It is worth noting that the baseline VS value also increases with cyclomatic complexity but it does so at a slower rate than any of the four LLMs. So, we can see that LLMs tend to struggle to produce good quality code when they are given highly complex tasks.  

\begin{figure}[htb]
\centering
\includegraphics[width=6.5cm]{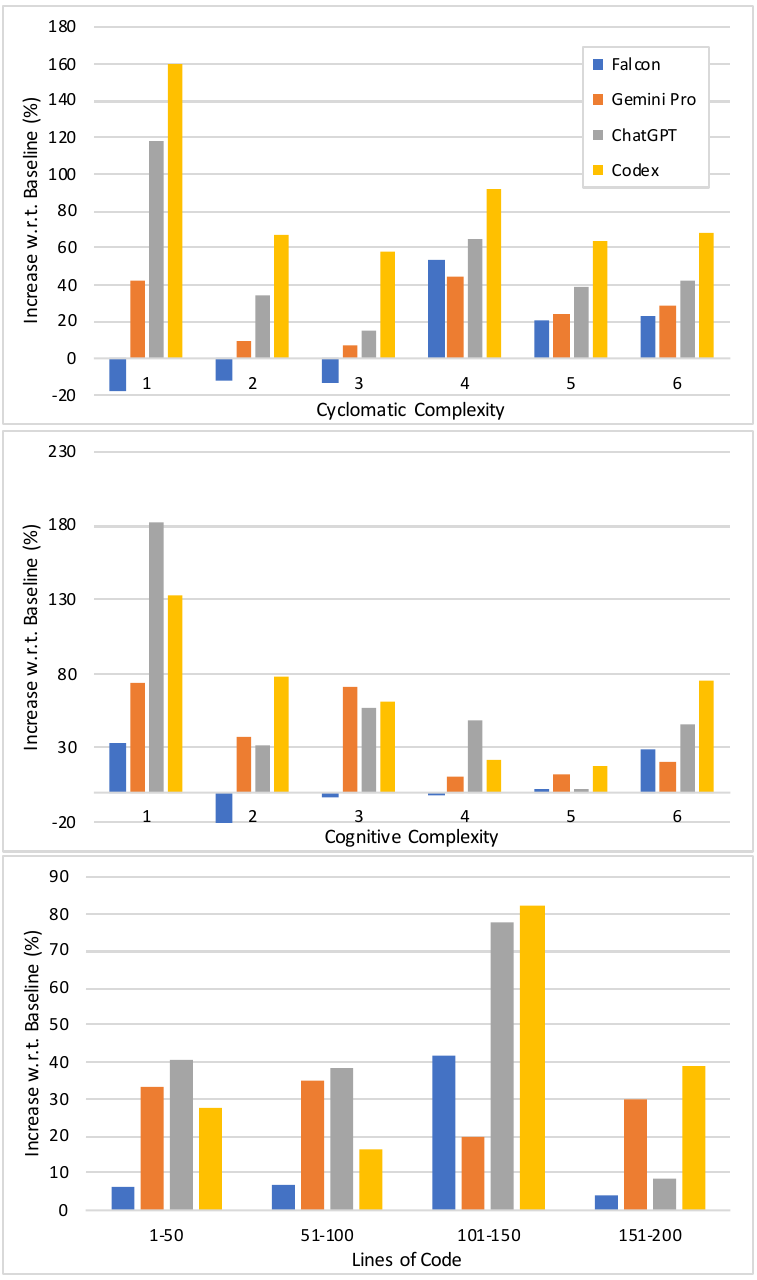}
\caption{Increase Rates of Code Smells by Complexity}
\label{fig:INcreaseOfByComplexity}
\end{figure}

However, by analysing the increase rate of code smells in the code generated by LLMs with respect to human-written  code,  we found no obvious link between the complexity of coding tasks and the increase rate of code smells as shown in Fig. \ref{fig:INcreaseOfByComplexity}. 

Therefore, we have the following observation. 

\begin{Observation} 
There is no clear evidence that the increase rate of code smells in LLM-generated code is correlated to the complexity of the coding task. 
\end{Observation}

\subsection{RQ4: Variation of Code Smells by Smell Types}

Research question \emph{RQ4} aims to identify the specific quality
issues in LLM-generated codes. We calculated the VS for each specific type of code smell on the whole test dataset and compared with the baseline. 

\begin{table*}[ht]
\caption{Code Smells by Smell Type}
\label{tab:code_smells_full}
\scriptsize
\centering
\includegraphics[width=18.5 cm]{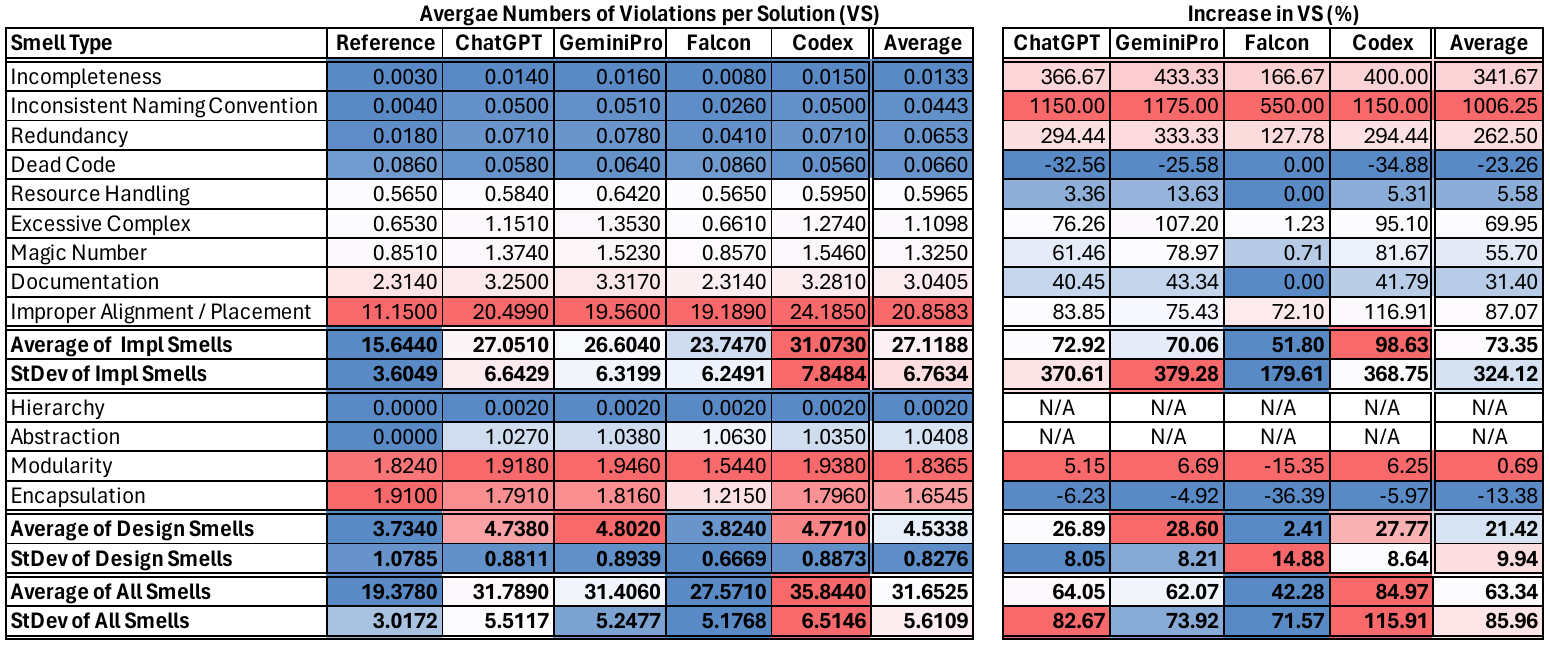}
\end{table*}

Table \ref{tab:code_smells_full} presents, in the form of a heat map, the VS for each specific type of smells as well as the increase rates for each LLM model in comparison with the baseline. The highest VS scores and the largest (i.e. worst) increase rate are highlighted in red, while the lowest VS scores and lowest  increase rates are coloured in blue. Implementation smells are listed in the top half and design smells in the bottom half. 

From the experiment data, we observed the following phenomena. 

\begin{Observation} 
The least prevalent types of implementation smells for all LLMs as well as the reference solutions are Incompleteness, Inconsistent Naming Convention and Redundancy. 
\end{Observation}

\begin{Observation} 
The worst types of implementation smells for all LLMs and human-written  code are Magic Number, Documentation and Improper Alignment and Placement. 
\end{Observation}

\begin{table}[htb]
\centering
\caption{Pearson Correlation Coefficients btw Baseline and LLMs' VS Scores on Various Smells Types}
\label{tab:VSTypeCorrelations}
\includegraphics[width=9cm]{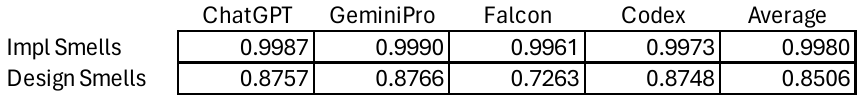} 
\end{table}

In general, there is a very strong correlation between LLM-generated code and human-written  code on the VS scores on the types of code smells. As shown in Table \ref{tab:VSTypeCorrelations}, for implementation types of code smells, the Pearson correlation coefficients between the baseline and the code generated by each LLM are all very close to 1. For design code smells, the Pearson correlation coefficients are in the range between 0.7263 for Falcon and 0.8766 for Gemini Pro. Therefore, we can confidently conclude that: 

\begin{Observation} 
The prevalence of code smells in LLM-generated code on various smell types is strongly correlated with that of human-written code. 
\end{Observation}

The experiment data also demonstrated that the prevalence of a type of code smell in human-written  code does not imply that the smell increases in LLM-generated code. As shown in Table \ref{tab:VS2IncreaseOnTypeCorrelations}, for implementation smells, the Pearson correlation coefficients between LLMs' smell increase rates and the VS scores of the baseline are all negative in the range between -0.2549 for Gemini Pro and -0.1518 for Falcon, where the average over all LLMs is -0.2188. 

\begin{table}[htb]
\centering
\caption{Pearson Correlation Coefficients btw Baseline VS Scores and LLMs' Increase Rates on Various Smells Types}
\label{tab:VS2IncreaseOnTypeCorrelations}
\includegraphics[width=9cm]{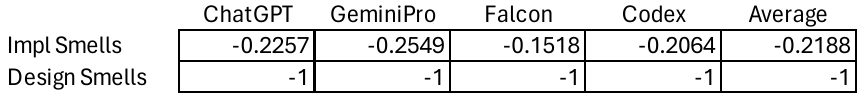}
\end{table}

However, it is observable that the largest increase rates of code smells are on the types that are the least prevalent smell types of the baseline. 

\begin{Observation} 
The largest increases of smells in LLM-generated code happened at the smell types of Incompleteness, Inconsistent Naming Convention and Redundancy.  
\end{Observation}

Finally, the Pearson correlation coefficients between LLM increase rates over various types of implementation smells are all very close to 1; see Table \ref{tab:IncreaseTypeCorrelations}. This implies the following observation. 

\begin{Observation} 
LLMs consistently increase the strength of code smells over various types of code smells. 
\end{Observation}

\begin{table}[htb]
\centering
\caption{Pearson Correlation Coefficients btw LLMs' Increase Rates of Implementation Smells}
\label{tab:IncreaseTypeCorrelations}
\includegraphics[width=8cm]{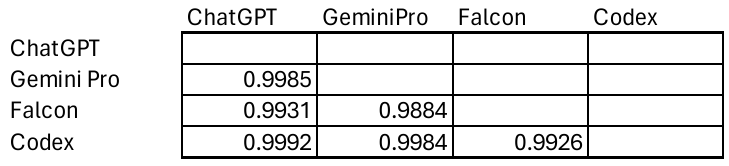}
\end{table}

From the experiment data, we can also have the following observations. 

\begin{Observation}
All LLMs performed well on the encapsulation type of design smells in comparison with the baseline. 
\end{Observation} 

\begin{Observation}
LLM performance on design smells vary significantly with increase rates ranging from 2.41\% for Falcon to 28.60\% for Gemini Pro, and the average increase rate over all LLMs is 21.42\%. Falcon performed better on design smells than the other LLMs. 
\end{Observation}

\begin{table}[htb]
\centering
\caption{The Most Prevalent Smells in Each Type}
\label{tab:TopSmells}
\includegraphics[width=9cm]{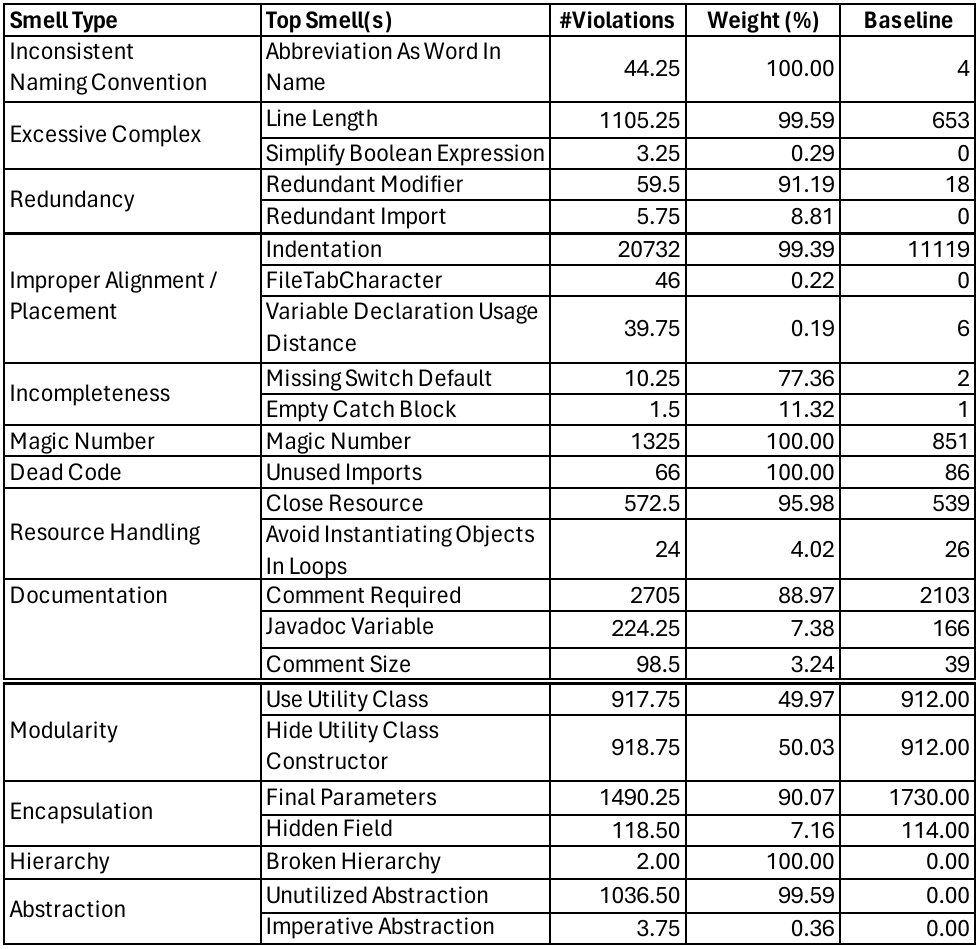}
\end{table}

Finally, by analysing the distributions of VS scores for each type of smells, we have the following observation. 

\begin{Observation}
For each type of smells, the violations of smell detection rules are concentrated in a small number of specific smells. 
\end{Observation}

Table \ref{tab:TopSmells} lists the most prevalent smells in each smell type, where column \emph{Top Smell(s)} lists the most prevalent smell(s) of the smell type given in the column \emph{Smell Type}.  Column \emph{\#Violations} gives the average numbers of violations of the specific smell rule over all LLMs. Column \emph{Weight} gives the ratio of the violations over all smells of the type. Column \emph{Baseline} gives the number of violations in the reference solutions.

Note that there are fewer violations of design smells than implementation smells. We believe that there are two reasons for this. First, the test dataset contains very few coding tasks where the solutions require a large number of classes.  In fact, only 68 (6.8\% of the dataset) require more than one class. Design smells like Hierarchy smells do not present in code that has only one class. Second, there are fewer detection rules for design smells than for implementation smells. However, fewer violations of design smells does not imply better design quality because design smells are at a higher level of abstraction and of greater granularity. Each violation of a design smell could have a more serious impact than one violation of an implementation smell. It is not meaningful to compare the number of violations of design smells to that of implementation smells. 

\subsection{RQ5: Variation of Code Smells by Correctness}

Research question \emph{RQ5} aims to understand how the correctness of LLM-generated codes relate to code smells. 

To determine the correctness of an LLM-generated solution, we employed the testing approach with $pass@1$ metric, which is the \emph{de facto} method widely used by researchers for evaluating LLMs on the correctness of generated programs \cite{ghosh2024b}. However, we enhanced the approach by generating test cases from both the reference solution and the LLM-generated code by employing the EvoSuite tool \cite{fraser2011evosuite,fraser2014large} rather than a fixed set of test cases associated with the coding task. These two sets of test cases are merged into one test suite and applied to both the reference solution and the generated code. This enables us to detect both omission errors and commission errors. In particular, if the reference solution fails on a test case that is generated from the LLM-generated code, a commission error is detected. If the generated code fails on a test case that is generated from the reference solution, an omission error is detected. The LLM-generated code is only regarded as correct by testing, i.e. it passes all of the tests, if neither a commission error nor an omission error are detected. In contrast, existing approaches that use a fixed set of test cases associated with each coding task can only detect omission errors. This approach has been proven to be able to achieve high test adequacy for LLM-generated code \cite{ghosh2024, Paul2025PhDThesis}. Readers are referred to \cite{ghosh2024,Paul2025PhDThesis} for technical details. 

\begin{figure}[htb]
\centering
\includegraphics[width=6.5cm]{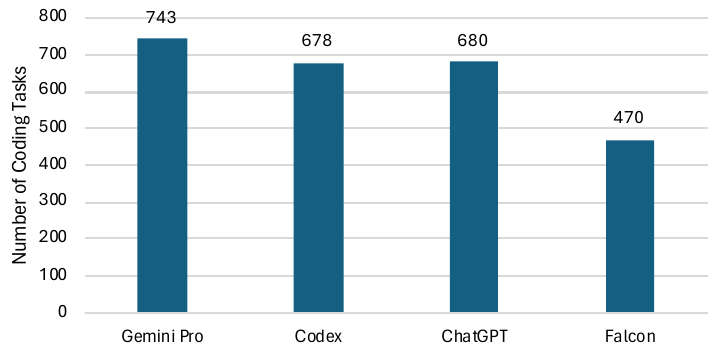}
\caption{Numbers of LLM-generated Solutions Passed Test}
\label{fig:PassRate}
\end{figure}

Fig. \ref{fig:PassRate} shows the number of LLM-generated programs that pass all tests. From the data shown in the figure, Gemini Pro performed the best with a success rate of 74.3\% passing all tests, while Falcon is the worst with a success rate only 47.0\%. Codex and ChatGPT performed very similarly, both have a success rate around 68.0\%. 

To analyse how correctness is related to code smell, we split the test dataset into two subsets: one contains coding tasks for which the LLM generated a correct code; the other contains tasks for which the LLM failed to generate a correct model. The average of smell violations per solution are calculated for each subset on all code smell detection rules. The results are shown in Table \ref{tab:SmellByCorrectness}. 

\begin{table}[htb]
\centering
\caption{Smells of Correct and Incorrect Codes}
\label{tab:SmellByCorrectness}
\resizebox{\linewidth}{!}{
\begin{tabular}{|l|c|c|c|c|c|c|}
\hline
\multirow{2}{*}{\textbf{LLM}}&\multicolumn{2}{|c|}{\textbf{Avg of Code Smells}}&\textbf{Increase}&\multicolumn{2}{|c|}{\textbf{Margin of Error (\%)}}&\textbf{p-value} \\\cline{2-3}\cline{5-6}
&\textbf{Correct} &\textbf{Incorrect}&\textbf{Rate (\%)}&\textbf{Correct} &\textbf{Incorrect}&\textbf{of T-Test}\\\hline
ChatGPT&31.3279&32.9344&5.1278&5.0862&6.5124&0.2391\\\hline
Gemini Pro&30.6299&33.8833&10.6216&5.0835&7.8972&0.0400\\\hline
Falcon&32.0511&23.6830&-26.1085&6.5659&6.2138&0.0000\\\hline
Codex&35.6416&36.4379&2.2342&4.5385&6.4342&0.5839\\\hline
\end{tabular}
}
\end{table}

From Table \ref{tab:SmellByCorrectness}, we can observe that correct code has fewer smells than incorrect code on average for three out of four LLMs: Gemini Pro, Codex and ChatGPT. However, Falcon is an exception. Its incorrect code has fewer smells than its correct code. Table \ref{tab:SmellByCorrectness} shows the rates of increase (\%) in code smells from correct code to the incorrect code for different LLMs. For Codex and ChatGPT, the increase rates are 2.23\% and 5.13\% respectively, indicating little difference in the code smells; while for Gemini Pro, the increase rate is to 10.62\%, indicating significant differences, and for Falcon, the increase rate of -26.11\% indicates a large amount of decrease in code smell. Therefore, we have the following observation. 

\begin{Observation}
The degree of differences between the correct and incorrect codes in terms of the strength of code smells varies significantly with the LLM models. For some LLMs, incorrect codes have more smells; while for the others, the opposite is observable. 
\end{Observation}

To verify the observation statistically, we have conducted a T-Test with the null hypothesis below.

\emph{H0: Given an LLM, there is no significant difference between the code smells in the programs generated by the LLM correctly and the code smells in the programs incorrectly generated by the LLM}. 

The p-values for various LLMs are also given in Table \ref{tab:SmellByCorrectness}. The results show that for Gemini Pro and Falcon are 0.0400 and less than 0.05 and 2.77179E-10 respectively, hence the null hypothesis can be rejected; while the p-values for Codex and ChatGPT are 0.2391 and 0.5839, far greater than 0.05, thus the null hypothesis cannot be rejected. 

It is worth noting that the average code smells per solution in both correct and incorrect subsets of test cases are of high confidence levels, as shown in the column \emph{Margin of Error (\%)} of Table \ref{tab:SmellByCorrectness}, the percentage margins of error of the average code smells are all below 10\% when the confidence level is 95\%. 

\section{Discussion on Threats to Validity}\label{sec:ThreatsToValidity}

In this section, we discuss the potential threats to the validity of the experiment reported in the previous section and how these threats were addressed in the design and conduct of the experiment. We will also discuss how to further reduce the threats in future work. We will apply Wohlin et al.'s framework \cite{wohlin2012experimentation} of classifying the threats to validity in software engineering experiments, as it is among the frameworks most used by the researchers in software engineering. 

\subsection{Construct Validity}

Construct validity is concerned with whether the data obtained by measurement and observation correctly and adequately represent the abstract concepts under study. In our context, it means whether the code smell detection rules correctly, adequately and fairly represent the quality aspects related to the readability, maintainability, ease of evolution, etc. Our experiments have included all smell detection rules implemented by multiple tools that we have used. Moreover, we have used more tools than existing works to combine the smell detection rules to achieve the maximal coverage of code smells. Therefore, we do not recognise any major threats to the construct validity from the coverage aspect. 

One potential threat to construct validity lies in the reliance on the accuracy of the static analysis tools used in our study. Any limitations or inaccuracies in these tools could affect the precision of our smell detection. To mitigate this threat, we selected widely used and validated tools (i.e., PMD, Checkstyle, and DesigniteJava) that have demonstrated reliability in prior research and practice. Moreover, it is reasonable to assume that these tools have the same error rate in detecting code smells in human-written code and LLM-generated code. By comparing the data between code smells of human-written and LLM-generated codes, the impact of the tool inaccuracy is limited. Thus, it does not impose a serious threat to the validity of the conclusions. 

Also, we focused solely on code smells detectable by PMD, Checkstyle, and DesigniteJava. While this may exclude certain types of smells, the selected set of detection rules represents a well-established and widely adopted list of code smells. These have been well documented and widely used both in academic research and industry practice, lending credibility to their relevance and maturity. Moreover, we have combined the smell detection rules provided by these tools to maximise the coverage of the smells. 

\subsection{External Validity}

The external validity is concerned with to what extent the results of an experiment can be generalised. A potential threat to external validity involves the generalisability of our findings to other LLMs not studied in our experiments, coding in programming languages other than Java, and those types of coding tasks not covered by the test dataset. 

Our analysis is based exclusively on Java programs from the ScenEval dataset, which may limit the applicability of the results to code generation tasks in other programming languages. 

Our experiments focus on the implementation and design smells. Therefore, the results are limited to implementation and design smells, and should not be generalised to architectural smells. 

Additionally, we evaluated outputs from LLMs GeminiPro, ChatGPT, Codex, and Falcon in Table \ref{tab:LLMs}. The landscape of LLMs evolves rapidly, and several more advanced code models have emerged since the completion of our experiments, including GPT-4/4o, Claude 3.5, Llama 3, Code Llama, StarCoder2, and DeepSeek-Coder. Consequently, the findings should be interpreted as a snapshot of the capabilities of leading models available during the study period rather than as a comprehensive assessment of the current state-of-the-art. However, there are observations that are consistent on all four LLMs that we studied. It would be interesting to study whether these observations could be generalised to other ML models. Future work should extend the evaluation to newer proprietary and open-source models to improve generalizability and reproducibility. 

Our dataset covers a wide range of topics. However, the majority of coding tasks are of small scale in terms of complexity. Moreover, very few of the coding tasks require multiple classes. The conclusions drawn from our experiment should be limited to the coding tasks well represented by our dataset. Any generalisation of our conclusions to other kinds of coding tasks should be used with great care. 

\subsection{Internal Validity}

Internal validity is concerned with the appropriateness of the design and conduct of the experiments. A typical example of the threats to internal validity is the existence of factors that influence the causal relationships under study but are not measured and are not under our control. 

A potential threat to the internal validity of our experiment is that LLMs are inherently nondeterministic. For this reason, we have selected a large number of test cases (1000) at random to minimise the impact of LLM's randomness. The scale of our experiment is much larger than most of the studies of LLMs' capability in code generation. For future research, this threat to internal validity can be further reduced by using even more test cases and repeating the invocations of LLMs on each coding task. 

Another potential threat to internal validity is that the quality of program code in general and code smells in particular are very subjective as we discussed in Section \ref{sec:Introduction} and \ref{sec:Background}. We have addressed this threat at the methodology level by excluding human factors from the experiment by using a baseline formed by professionally written code and at the technology level by employing the quantitative analysis of the experiment data using objective metrics. 
 
Finally, a potential threat to internal validity is that the implementation of the test system may contain bugs, thus the data collected may have errors. We have addressed this threat by careful testing and debugging of the test system. Moreover, to ensure the experimental reproducibility, the source code of the test system as well as the data are available to the public in the GitHub repository \footnote{URL: https://github.com/hongzhu6129/EvaluateLLMCodeSmell}. 

\subsection{Conclusion Validity}

Conclusion validity is concerned with whether the conclusion drawn from the experiment is logically valid, such as whether correct statistical inference methods are used properly and whether the statistical inference power is strong enough. 

In the research on RQ1, we calculated the margins of error of the statistical data about the average number of violations of code smell per solution at the confidence level of 95\%. The results show that  the margins of errors are in the range between 3.708\% and 6.045\%. Moreover, we have also conducted t-tests on the significance of the differences between VS values of LLM-generated code and VS value of human-written  code. The p-values are in the range between 3.938E-68 and 3.0557E-42. Therefore, the observations made for answering RQ1 are at a high confidence level.  

In the research on RQ2, RQ3 and RQ4, the relationships between the variations of code smells with code topic, coding task complexity and smell types in human-written and LLM-generated code have been statistically analysed through regression analysis and Pearson correlation coefficients were calculated. The observations are drawn based on the regression analysis results with Pearson correlation coefficients very close to 1. Thus, the observations are of high statistical inference strengths. 

In the research on RQ5, we also analysed the margin of errors in the VS values for correct and incorrect codes generated by LLMs with confidence level 95\%. The margins of error are in the range between 4.539\% and 7.897\%. We have also conducted t-tests on the significance of the differences between the VS values of correct and incorrect codes. The observations are drawn based on results of the t-tests (i.e. the p-values). 

Therefore, the statistical analysis and inference methods are valid and appropriate and the conclusions are logically valid and of high confidence.  

\section{Conclusion and Future Work} \label{sec:Conclusion}

In this paper, we proposed a scenario-based methodology to evaluate the usability of LLM-generated code on readability, modifiability, reusability, maintainability, easiness of evolution, etc., through assessing the code smells and comparing with a baseline obtained from code written by professional programmers. An automated test system is designed and implemented in the datamorphic testing method. An intensive experiment with four prominent LLMs is conducted using the ScenEval benchmark for generating Java code. 

We have found that the code smells, measured by the average number of violations of code smell detection rules per solution, is significantly greater in LLM-generated code compared to human-written code. Across all LLMs, the average increase rates of
implementation and design smells are 73.35\% and 21.42\%, respectively, while the average increase rate over all smells is 63.34\%. 

The performances of LLMs vary significantly over different topics of coding tasks and smell types. In general, the more complicated a coding task is, the stronger is the smell in LLM-generated code. The types of code smells that are the strongest in human-written  code are also the most prevalent in LLM-generated code. However, the increase rates of code smell types in LLM-generated code show no correlation to the prevalence of the type of code smell in human-written  code. In general, the quality of generated code decreases with the complexity of the code task.  This correlation is very clear when coding task complexity is measured by cyclomatic complexity and the lines of code, but is slightly less clear with cognitive complexity. 

For future work, it is worth further expanding and repeating the experiments with more LLM models and using a larger test dataset to reduce the risks of the potential threats to validity as discussed in Section \ref{sec:ThreatsToValidity}. 

As discussed in Section \ref{sec:Background}, it is difficult to set a threshold on the number of violations for the code 
to be of acceptable quality. We could only compare with the baseline number,
which reflects current practice. Hence, from our experiment data, we have difficulty to answer the
question: is the quality of LLM-generated code acceptable for use? 
We encourage further research in this area. 

It is worth noting that the number of code smell violations in a piece of code depends on the settings of the thresholds for the detection rules of the tools. The thresholds in the smell detection rules that we used in the experiments are the default values of the tools, which are the best settings based on the results of their research. They have been widely used by researchers in the subject domain and this aligns well with the related works so that our results are compatible with other existing works. The sensitivity of these thresholds for LLM-generated code are worth an intensive and careful study through experiments.

An implicit assumption underlying our study of the code smells in LLM-generated code and the comparison with human-written code is that the smell detection techniques in general and the tools in particular work equally well on both types of code. Since the code smell detection techniques and the tools are based on research on human-written  programs, a research question that remains open is whether LLM-generated codes have any special type of smells that could prevent the code from being used but cannot be detected by existing code smell tools and techniques. Therefore, a manual validation of the code smells detected by the tools and manual inspection of the code smells of LLM-generated code are worth performing as a future work.

It is interesting to know if there are significant differences between open-source LLMs and proprietary LLMs in the code smells of generated programs. In our experiments, Falcon is the only open-source LLM and its behaviour seems different from the proprietary LLMs. The under-representation of open-source LLMs in our experiments prevents us from drawing a general conclusion. Therefore, it is worth further study on whether this phenomenon can be generalised to other open-source LLMs.

In this paper, we have focused on implementation and design smells. Architectural smell concerned with architectural design issues that are presented in program code. Since LLMs have not been widely used in practice for architectural level code generation and there is no well-developed benchmark of such coding tasks with reference solutions, we cannot assess architectural smell reliably and efficiently. It is an open problem for future research.

Moreover, LLM-generated code has broader quality aspects of usability that are worth considering, such as security and runtime efficiency of the generated code, which is addressed by a recent work by Jonnala et al. \cite{jonnala2025}. This way we hope to provide a more holistic
evaluation of LLM performance for code generation.   

It is also worth investigating how code smell detection can be used to improve the quality of LLM-generated code, for example, in a multi-attempt process proposed by Miah and Zhu \cite{miah2024user}. We have already conducted experiments in the design and implementation of a multi-agent system in which an agent detects the smell of the code generated by an LLM-based agent. The detected code smells are then fed to another agent for reducing the code smell using LLM \cite{Paul2025PhDThesis}. Due to the limitation of space, the work will be reported in a separate paper. 

Another direction of research worth pursuing is to fine tune LLMs or train new LLMs with code dataset in which code smells are reduced and cleaned. The works by Siddiq et al. in 2022 \cite{siddiq2022empirical} and Xue et al.  more recently\cite{xue2025clean} suggest that code quality could be improved by using such training data for fine-tuning of code models. Our results that code smells in LLM-generated codes highly correlate to the code smell in human-written code also suggest that the smells in LLM-generated code are caused by smells in the training data. 

\section*{Acknowledgement}
The research work reported in this paper is partly funded by Google's PaliGemma
Academic Program, which provided credits for using Google Cloud Platform resources.  

\bibliographystyle{elsarticle-num}
\bibliography{references} 

\end{document}